\documentclass[final,3p,times,authoryear]{elsarticle}

\usepackage{hyperref}

\usepackage[T1]{fontenc}
\usepackage[utf8]{inputenc}
\usepackage{lmodern}
\usepackage{amssymb,amsmath,amsthm,textcomp}
\usepackage{graphicx}
\usepackage{wrapfig}
\usepackage[english]{babel}
\usepackage{csquotes}
\usepackage{paralist}

\usepackage{color,ifthen}
\usepackage{xcolor}

\usepackage{fontawesome}
\usepackage{marvosym}

\usepackage{commentingCommands}
\newcounter{contextexamp}[section] 
\renewcommand{\thecontextexamp}{C.\arabic{contextexamp}}

\newcommand{\startingPoint}[1]{\vspace{1ex}\noindent
{\large\faMapMarker\hspace{1.5ex}}%
\ifthenelse{\equal{#1}{}}{\textbf{One starting point}}{#1}%
}

\newcounter{goalPL}
\renewcommand{\thegoalPL}{G.\arabic{goalPL}}
\newenvironment{goal}[1][]{
\refstepcounter{goalPL}
\par\vspace{1ex}\noindent%
\textbf{\thegoalPL}\ifthenelse{\equal{#1}{}}{\textbf{:}}{\textit{\hspace{1ex}(#1):}}%
\begin{itshape}
}{
\end{itshape}\hfill\faFlagCheckered\vspace{1ex}
} 
\newcommand{\refGoal}[1]{\ref{#1}}

\newcounter{characterPL}
\renewcommand{\thecharacterPL}{C.\arabic{characterPL}}
\newenvironment{characteristic}[1][]{
\refstepcounter{characterPL}
\par\vspace{1ex}\noindent%
\textbf{\thecharacterPL}\ifthenelse{\equal{#1}{}}{\textbf{:}}{\textit{\hspace{1ex}(#1):}}%
\begin{itshape}
}{
\end{itshape}\hfill\faDiamond\vspace{1ex}
} 
\newcommand{\refCharacteristic}[1]{\ref{#1}}

\newcounter{openprobPL}
\renewcommand{\theopenprobPL}{OP.\arabic{openprobPL}}
\newenvironment{openproblem}[1][]{
\refstepcounter{openprobPL}
\par\vspace{1ex}\noindent%
\textbf{\theopenprobPL}\ifthenelse{\equal{#1}{}}{\textbf{:}}{\textit{\hspace{1ex}(#1):}}%
\begin{itshape}
}{
\end{itshape}\hfill\faGraduationCap\vspace{1ex}
} 
\newcommand{\refOpenProblem}[1]{\ref{#1}}

\newcommand{\MultyLong}{\textit{Multy:~}}
\newcommand{\Bussy}{\textit{B:~}}
\newcommand{\Lancey}{\textit{L:~}}
\newcommand{\Reggy}{\textit{R:~}}
\newcommand{\Upsy}{\textit{U:~}}
\newcommand{\Eddy}{\textit{E:~}}
\newcommand{\Techy}{\textit{T:~}}
\newcommand{\Multy}{\textit{M:~}}
\renewcommand{\Bussy}{\textit{Bussy:~}}
\renewcommand{\Lancey}{\textit{Lancey:~}}
\renewcommand{\Reggy}{\textit{Reggy:~}}
\renewcommand{\Upsy}{\textit{Upsy:~}}
\renewcommand{\Eddy}{\textit{Eddy:~}}
\renewcommand{\Techy}{\textit{Techy:~}}
\renewcommand{\Multy}{\textit{Multy:~}}
\newcommand{\BussyName}{\textit{Bussy}}
\newcommand{\LanceyName}{\textit{Lancey}}
\newcommand{\ReggyName}{\textit{Reggy}}
\newcommand{\UpsyName}{\textit{Upsy}}
\newcommand{\EddyName}{\textit{Eddy}}
\newcommand{\TechyName}{\textit{Techy}}

\longPaper{
\journal{A Journal}
}\onlyShortPaper{%
\journal{Computer Law \& Security Review}
}

\begin{document}

\begin{frontmatter}

\longPaper{%
\title{A Multidisciplinary Definition of Privacy Labels: The Story of Princess Privacy and the Seven Helpers\tnoteref{labelThanks}%
}

\tnotetext[labelThanks]{We would like to thank associate Torunn Hellvik Olsen for her great inputs during our workshop on this topic held in Oslo, March 2020.}

}\onlyShortPaper{%
\title{A Multidisciplinary Definition of Privacy Labels: The Story of Princess Privacy and the Seven Helpers%
}
}

\anonymousPaper{
\author{Authors Anonymous}
}
\notAnonymousPaper{
\author[uio]{Johanna Johansen\corref{cor1}}
\ead{johanna@johansenresearch.info}
\cortext[cor1]{Corresponding author's address: P.O.box 1080 Blindern, 0316 Oslo, Norway. E-mail: johanna@johansenresearch.info }
\address[uio]{Dept.\ of Informatics, University of Oslo}
\author[bjorknes]{Tore Pedersen}
\ead{tore.pedersen@bhioslo.no}
\address[bjorknes]{Bj\o{}rknes University College}
\author[karlstad]{Simone Fischer-H\"{u}bner}
\ead{simone.fischer-huebner@kau.se}
\address[karlstad]{Dept.\ of Mathematics and Computer Science, Karlstad University}
\author[ntnu]{Christian Johansen}
\ead{christian.johansen@ntnu.no}
\address[ntnu]{Norwegian University of Science and Technology}
\author[chalmers]{Gerardo Schneider}
\ead{gerardo@cse.gu.se}
\address[chalmers]{Dept.\ of Computer Science and Engineering, University of Gothenburg}
\author[privacy]{Arnold Roosendaal}
\ead{arnold.roosendaal@privacycompany.nl}
\address[privacy]{Privacy Company}
\author[uld]{Harald Zwingelberg}
\ead{hzwingelberg@datenschutzzentrum.de}
\address[uld]{Unabh\"{a}ngiges Landeszentrum f\"{u}r Datenschutz Schleswig-Holstein}
\author[uio]{Anders Jakob Sivesind}
\ead{ajsivesind@gmail.com}
\author[its]{Josef Noll}
\ead{josef.noll@its.uio.no}
\address[its]{Dept.\ of Technology Systems, University of Oslo}
}

\begin{abstract}

Privacy is currently in distress and in need of rescue, much like princesses in the all-familiar fairytales. 
We employ storytelling and metaphors from fairytales to make reader-friendly and streamline our arguments about how 
a complex 
concept of Privacy Labeling (the `knight in shining armor') can 
be a solution to the current state 
of Privacy (the `princess in distress').
We give a precise definition of Privacy Labeling (PL), 
painting a panoptic portrait 
from seven different perspectives (the `seven helpers'): Business, Legal, 
Regulatory, 
Usability and Human Factors, Educative, Technological, and Multidisciplinary. 
We describe a common vision, proposing 
several important 
`traits of character' of PL as well as identifying 
`undeveloped potentialities', 
i.e., open problems on which the community can focus.
More specifically, this position paper identifies the stakeholders of the PL and their needs with regard to privacy, describing how PL should be and look like in order to address these needs.
Throughout the paper, we highlight goals, characteristics, open problems, and starting points for creating, what we consider to be, the ideal PL. In the end we present three approaches to establish and manage PL, through: self-evaluations, certifications, or community endeavors. Based on these, we sketch a roadmap for future developments.
\end{abstract}


\end{frontmatter}

\section{Introduction}

\notAnonymousPaper{
\onlyShortPaper{
\footnotetext{A long version of this paper exists at \url{https://arxiv.org/abs/2012.01813} .}%
}
}

The right to privacy is something precious and frail (an integral value appearing in the Universal Declaration of Human Rights\footnote{The ``right to privacy'' emerged in the Universal Declaration of Human Rights, adopted in 1948, as one of the fundamental human rights.%
\longPaper{ Shortly after, this right was reaffirmed in the European Convention on Human Rights (ECHR), drafted in 1950.}}), which we need to take good care of in order not to lose it, much like princesses in fairytales.
Just like European royalties, privacy is known to people only as a symbol, but does not have much power in the economy or society. In its current state it does not always serve the people, but mainly a few very wealthy and influential actors prosper from its misuse.
Loss of privacy has both micro implications, at a personal level (e.g., people being influenced to buy what they do not want or need \citep{Matz12714}, to vote for extremists \citep{2018CamAnalytica,berghel2018malice,stewart2019information}, or to develop antisocial behavior), but also macro implications, at a societal level (e.g., a society living in fear of being watched by surveillance capitalists \citep{zuboff2019age} or manipulated on social media \citep{starbird2019disinformation,grinberg2019fake}).
Privacy is personal and contextual, having social and political ramifications, but most of the population does not see, or understand, even some of its basic implications.
The lack of privacy literacy can be partly attributed to commercial entities that often, while profiting from handling data, work hard to keep privacy ``out-of-sight [is out-of-mind]'' -- like a sleeping princess locked in a tower -- e.g., telling people infamously ``You have zero privacy anyway. Get over it.''\footnote{%
\url{https://www.wired.com/1999/01/sun-on-privacy-get-over-it/}} \citep{solove2011nothing}.
Privacy misapprehension by the population is also due to its complexity, having kept many brilliant minds preoccupied for at least a century, since photography as a new technology used by media became widespread \citep{brandeis1890right}. In the current digital society, privacy \citep{solove2004digital,acquisti2007digital}
has even stronger forces compounding its complexity, coming from, e.g., technological advances in miniaturizing hardware that enabled cheep privacy-invasive gadgets, powerful algorithms that can make inconceivable inferences \citep{schneier2015data}, or supercomputing in `invisible' clouds \citep{borning2020invisible}; all too complex for laypeople to grasp.
It is fair to say that against such rapidly changing technologies, a person alone, no matter how dedicated she may be, would find it impossible to protect her Princess Privacy.

\anonymousPaper{
\begin{figure*}[t]
\centering
\includegraphics[width=11cm]{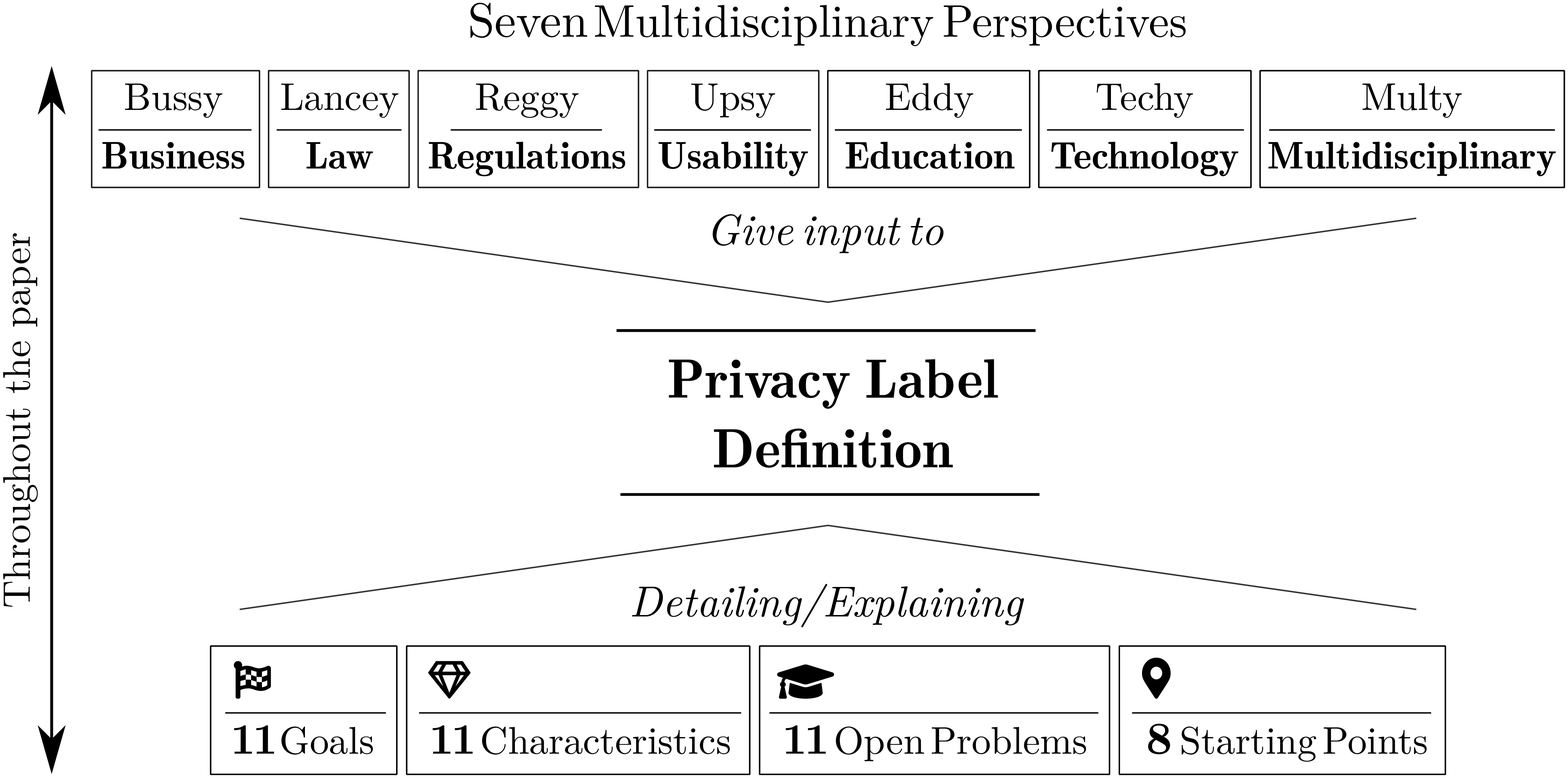}
\caption{Diagrammatic summary of contributions.}\label{fig_contributions}
\end{figure*}
}

This paper analyzes how the concept of Privacy Labeling/Labels, which hereafter we refer to as PL, could contribute to resolving several of the challenges that privacy is currently facing.
Given the many facets of privacy and its society-wide implications, it is important to adopt a multidisciplinary approach. 
This work started from a workshop in spring 2020 where experts from different fields of practice and research gathered to present and discuss their views on the topic of PL.
We thus bring in 
the following perspectives:
\longPaper{
\begin{itemize}
\item Business (relevant topics including, e.g.,
market potential, incentives, social responsibility, added value);
\item Law (e.g., compliance, privacy policies, audit);
\item Regulations (e.g., national, European, implementations, domain-specific standards); 
\item Usability and human factors (e.g., personas, easy to understand, completeness, contextual);
\item Education (e.g., psychology of people, of SMEs (Small and Medium-sized Enterprises), of CEOs (Chief Executive Officers), nudging for good, mental heuristics); 
\item Technology (e.g., AI, reasoning, automation, dynamic labels, verification); 
\item Multidisciplinary (e.g., communication across fields, people, or companies, making synergies).
\end{itemize}}%
\onlyShortPaper{%
\textit{Business} (relevant topics including, e.g.,
market potential, incentives, social responsibility, added value);
\textit{Law} (e.g., compliance, privacy policies, audit);
\textit{Regulations} (e.g., national, European, implementations, domain-specific standards); 
\textit{Usability} and human factors (e.g., personas, easy to understand, completeness, contextual);
\textit{Education} (e.g., psychology of people, of SMEs (Small and Medium-sized Enterprises), of CEOs (Chief Executive Officers), nudging for good, mental heuristics); 
\textit{Technology} (e.g., AI, reasoning, automation, dynamic labels, verification); 
\textit{Multidisciplinary} (e.g., communication across fields, people, or companies, making synergies).}

We elaborate on
how to combine the seven different perspectives, the roles and priorities of each of these in relation to PL,  and point to the state of affairs in the respective fields.

Since it is rather intricate to provide a completely comprehensive picture of PL, we chose a storytelling style of discourse, and use the story of “Snow White and the Seven Dwarfs” as our parable.
This inspired us to 
employ metaphors such as the `seven helpers' as an analogy for our seven perspectives and `Princess Privacy' as the one to be saved by the `Privacy Labeling Knight'. 
We give each helper a name, and we use it to mark parts of the text with the respective perspective it belongs to:
\longPaper{
\begin{itemize}
\item \textit{Bussy} -- bringing in business arguments,%
\item \textit{Lancey} -- bringing in the legal perspective/argument,
\item \textit{Reggy} -- bringing in the regulatory perspective/argument,
\item \textit{Upsy} -- bringing in the usability perspective/argument,
\item \textit{Eddy} -- bringing in the educational perspective/argument,
\item \textit{Techy} -- bringing in the technological perspective/argument,
\item \textit{Multy} -- bringing in the perspective/argument of multidisciplinarity.
\end{itemize}
}%
\onlyShortPaper{%
\textit{Bussy} -- bringing in business arguments,
\textit{Lancey},
\textit{Reggy},
\textit{Upsy},
\textit{Eddy},
\textit{Techy}, and
\textit{Multy} respectively.}

We define the concept of \emph{Privacy Labeling}\footnote{A 90 seconds `elevator-pitch' video where we present Privacy Labeling for a general public can be viewed online at: \url{https://youtu.be/noE_vF2_GEs}.} below, and throughout the paper we detail each of its elements.

\begin{itemize}
\item[] \textit{A Privacy Label is a legally binding label containing information about the privacy that a product or service provides. The labels may be physical or digital. They are defined, and are visually presented, in a layered manner, where one can drill-down from a simple overview to more complex information, to allow the user to  
focus on different levels of detail, depending on the intended use.
The labels measure privacy using graded-scales to make it easy to compare two labeled products with respect to privacy aspects relevant for a particular (type of) user.}
\end{itemize}

More specifically, one can imagine PL as being similar to both nutrition facts labels and energy consumption labels. To make PL legally binding one can tie it to a privacy policy text, so that it cannot become a means of deceit in the hands of product advertisers, thus going beyond, but not against, laws and regulations such as GDPR.\footnote{The European General Data Protection Regulation (GDPR) \citep{EU2016GDPR}.}
PL are promoting ``privacy as an added value'' to a digital product,\longPaper{\footnote{%
\url{https://www.nbcnews.com/tech/security/can-privacy-be-big-business-wave-startups-thinks-so-n1128626}}$^{,}$\footnote{ENISA. ``Study on monetising privacy. An economic model for pricing personal information''. \url{https://www.enisa.europa.eu/publications/monetising-privacy}}} allowing privacy conscious businesses to differentiate themselves from those market competitors that prefer to monetize on the big-data model at the expense of the privacy of the user\longPaper{\footnote{ENISA. ``The Value of Personal Online Data''. \url{ https://www.enisa.europa.eu/publications/info-notes/the-value-of-personal-online-data}}}.
\onlyShortPaper{
As such, Privacy Labels should
(i) be educational (``Oh, there's a notion of privacy for this TV-set!''), 
(ii) be an incentive and promote business differentiation (business slogans could sound like: ``We care about your privacy. So should you!''), 
(iii) be legally conscious yet 
(iv) usable for the layperson 
(``Hey son, what’s all this writing about privacy here?''), 
hence with sufficient detail as needed, 
yet visual and simple, 
(v) be taken up into regulations and 
(vi) supported by technologically innovative tools.
Including all these characteristics requires a multidisciplinary effort.
}\longPaper{
As such, Privacy Labels should:
\begin{enumerate}
\renewcommand{\theenumi}{\roman{enumi}}
\renewcommand{\labelenumi}{(\theenumi)}
\item be educational (``Oh, there's a notion of privacy for this TV-set!''), 
\item be an incentive and promote business differentiation (business slogans could sound like: ``We care about your privacy. So should you!''), 
\item be legally conscious yet 
\item be usable for the layperson 
(``Hey son, what’s all this writing about privacy here?''), 
hence with sufficient detail as needed, 
yet visual and simple, 
\item be taken up into regulations and 
\item supported by technologically innovative tools.
\end{enumerate}
Including all these characteristics requires a multidisciplinary effort.
}

\notAnonymousPaper{
\begin{figure*}[t]
\centering
\includegraphics[width=13cm]{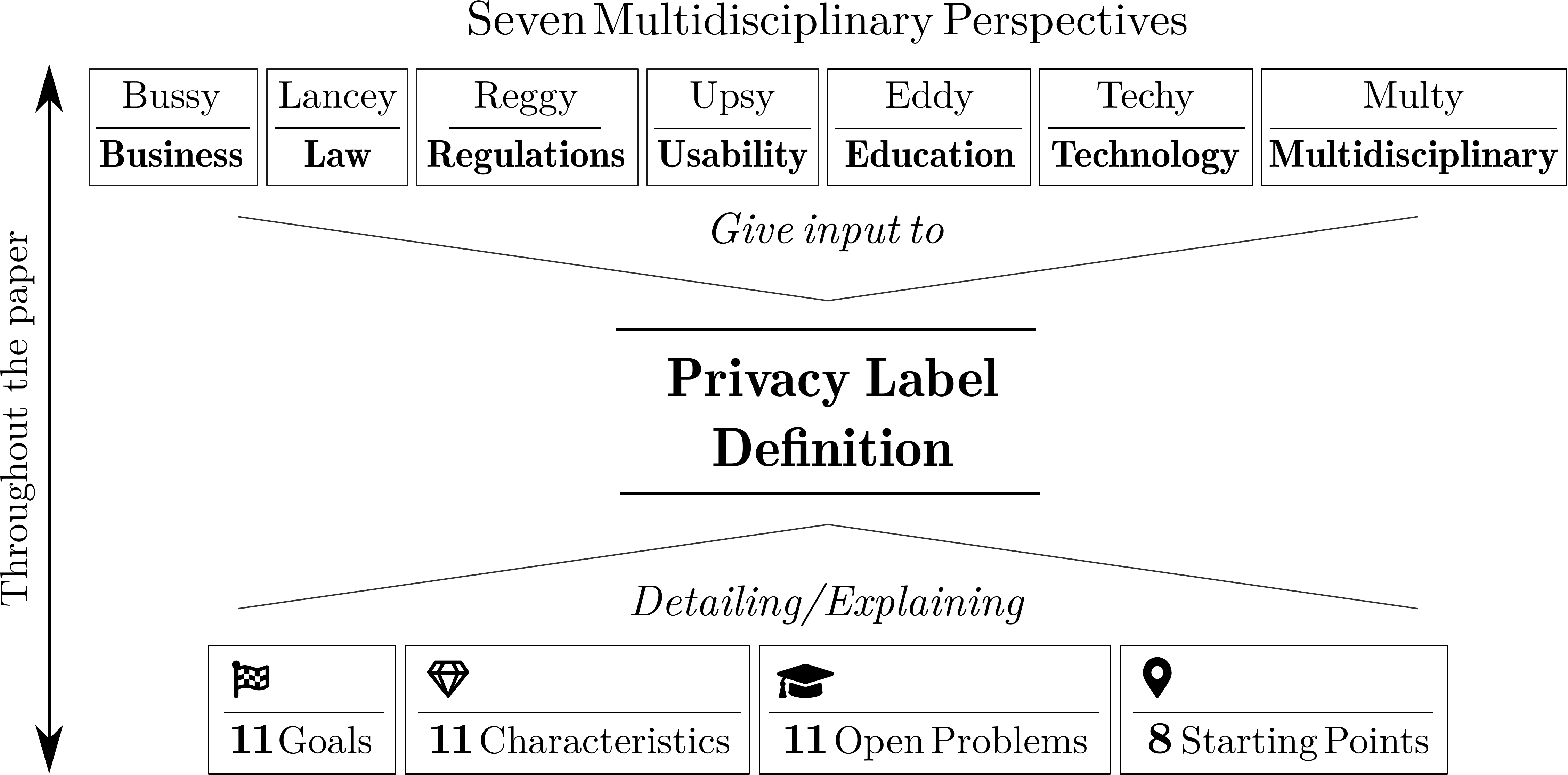}
\caption{Diagrammatic summary of contributions.}\label{fig_contributions}
\end{figure*}
}

\anonymousPaper{
\begin{figure*}[t]
\centering
\includegraphics[width=14cm]{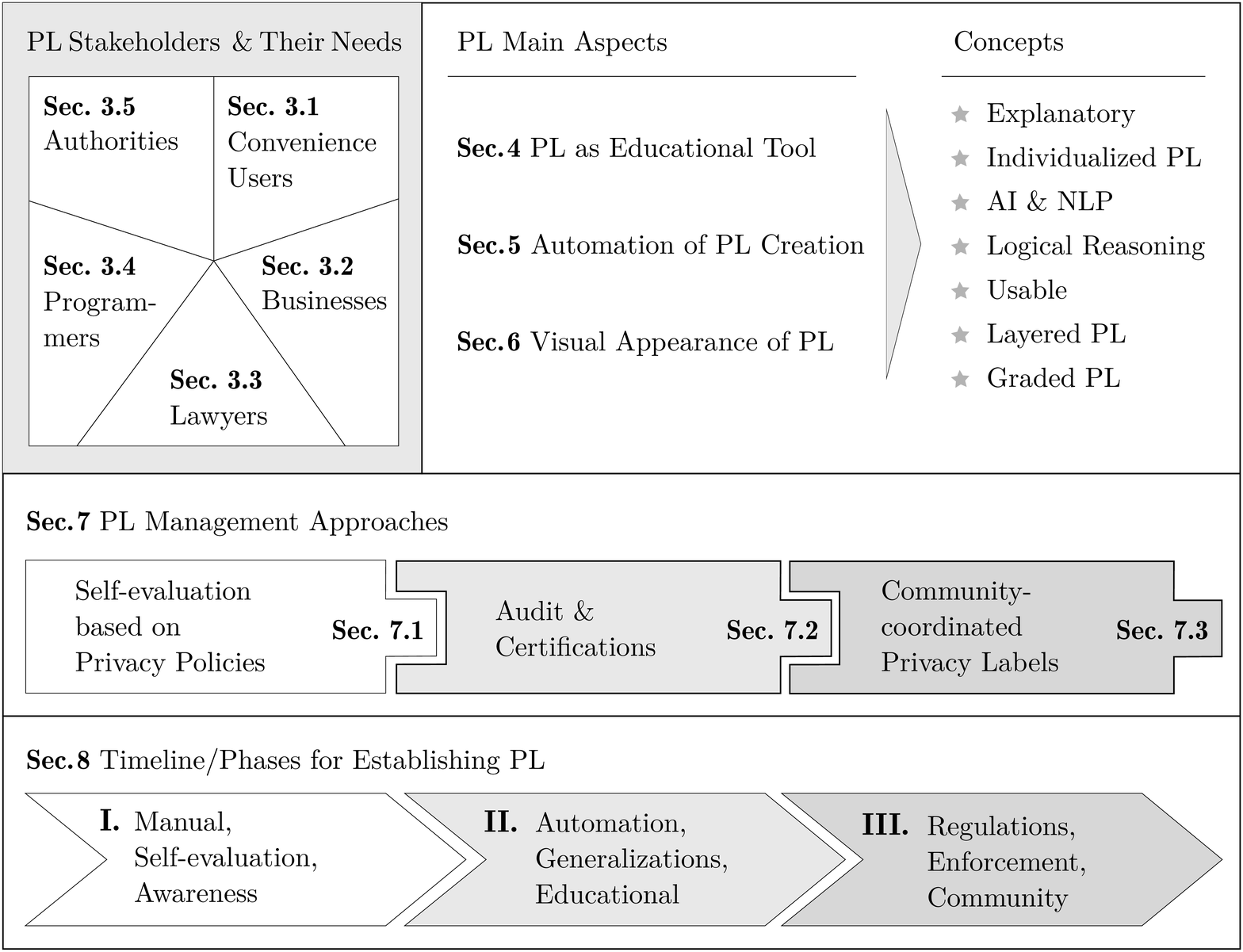}
\caption{Overview of the Paper.}\label{fig_structure}
\end{figure*}
}

\vspace{1ex}
Apart from defining PL and discussing it from all the seven relevant viewpoints, we propose along the way: goals, characteristics, open problems, and starting points for research. 
Further \emph{contributions} can be summarized as follows (and are schematically presented in Figure~\ref{fig_contributions}).

\begin{compactitem}
 \item We identify how the present landscape of privacy certifications (including privacy seals and marks \citep{rodrigues2018privacy}) could be improved by PL (see Section~\ref{sec_sleeping_princess_privacy}). 
\item We identify the stakeholders of PL, what are their needs and characteristics, as well as the relation between them (see Section~\ref{sec_PL_Stakeholders}).
\item Three important aspects of PL are presented: 
in Section~\ref{sec_PL_for_Edu}, its educational power to change people's knowledge of privacy;
in Section~\ref{sec_automation}, tools useful for constructing PL;
in Section~\ref{sec_looks_and_appearance}, the possible visual appearances of PL.
\item Three approaches to obtaining PL are presented in Section~\ref{sec_approaches} and a roadmap for achieving PL is outlined in Section~\ref{sec_roadmap}. 
\end{compactitem}

The above listed contributions constitute the major points of debate for the seven helpers. Imagine them sitting around the sleeping Princess Privacy discussing how to find a `Privacy Labeling Knight' with the  traits of character needed to awaken the princess. As the stories often depict it, not any knight will be right for the task.
Therefore, throughout the paper we point out: 
\begin{itemize}
\item goals (numbered as \textbf{G.x} and marked with a flag icon \faFlagCheckered ) 
intended to be achieved with the help of PL;

\item characteristics that we think PL should have (numbered as \textbf{C.x} and marked with a diamond icon \faDiamond );

\item open problems (numbered as \textbf{OP.x} and marked with a scholar cap icon \faGraduationCap ) 
that the community can address while striving for reaching any of the above;

\item existing works that can function as good starting points (marked with a map-pointer icon \faMapMarker )  for some of the above.
\end{itemize}
Following the practice of ``eating our own dog food'',
we mark the identified goals, characteristics, and open problems, with icons to make these important contributions of this paper more accessible (see in Section~\ref{sec_looks_and_appearance} our discussions on the use of icons in PL).

We believe that collecting all the {\faDiamond}/characteristics would form an ideal PL that could be useful to attaining the {\faFlagCheckered}/goals that we have pointed out. 
To help the community work towards creating such a PL we outline several {\faGraduationCap}/open problems and also identify good {\faMapMarker}/starting points among the existing works, along with drawing, in the end of the paper, a general roadmap to follow while taking one (or more) of the three approaches that we propose for managing PL.
\longPaper{Since all of these are the results of the dialogue and agreements between the seven helpers' different viewpoints, we expect the acceptance and usefulness of PL within the society to be considerable.}

In the next section, the seven helpers examine why Princess Privacy is asleep, what is the cause of this present dark situation 
and how to make the future brighter for their princess by describing the impacts and benefits of PL 
detailed in the rest of the paper.
The structure of this paper is displayed in Figure~\ref{fig_structure}.
An example that we use throughout the paper is the \emph{PrivacyLabel.org}\footnote{\url{https://www.privacylabel.org/} project from the Netherlands.}, which we hereafter refer to as NL.PL.

\notAnonymousPaper{
\begin{figure*}[t]
\centering
\includegraphics[width=15cm]{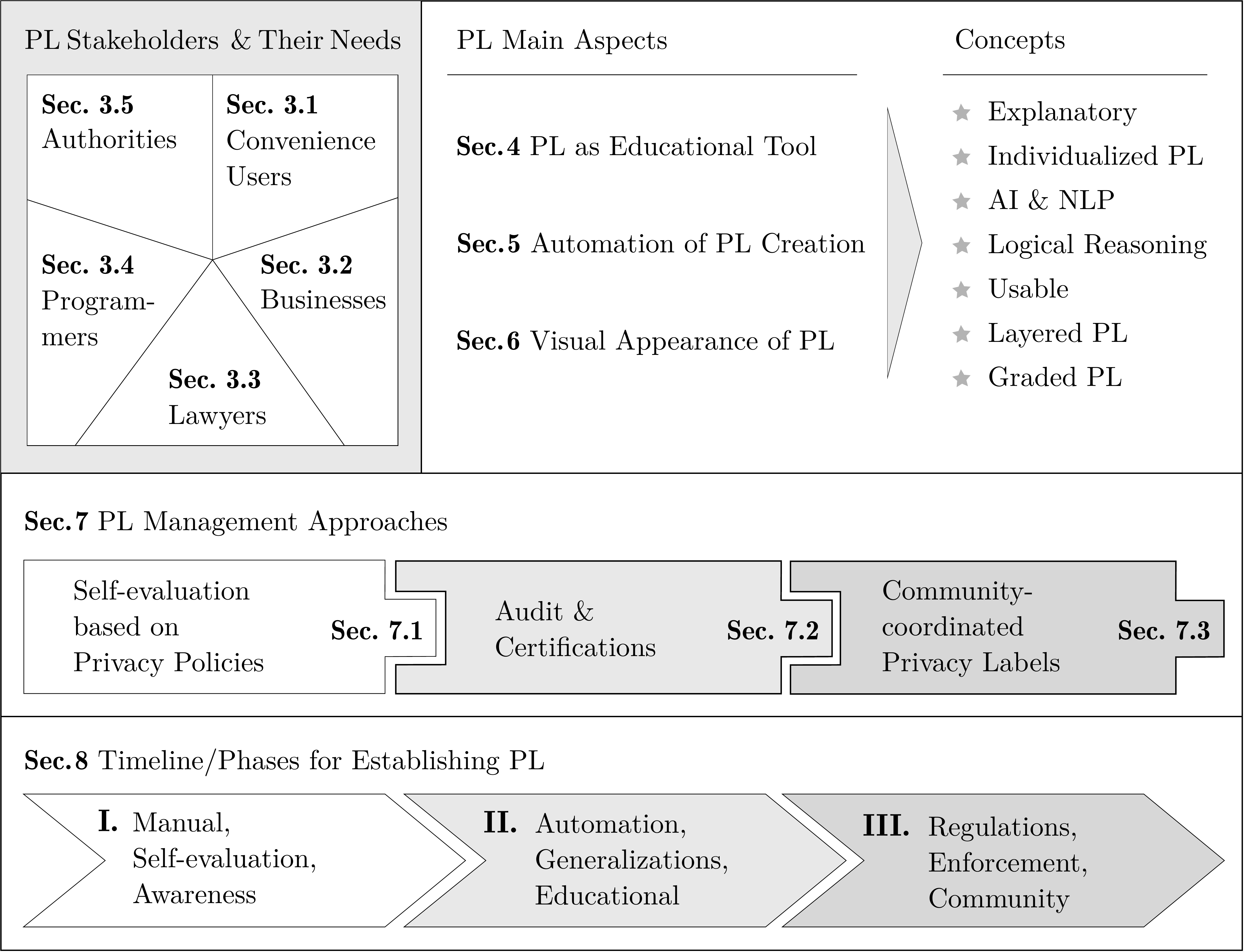}
\caption{Overview of the Paper.}\label{fig_structure}
\end{figure*}
}

\section{Sleeping Princess Privacy and the Privacy Labeling Knight}\label{sec_sleeping_princess_privacy}\onlyShortPaper{\label{sec_evaluating_situation}}

\MultyLong%
If Privacy Labeling is to be the `knight savior' we need a multidisciplinary effort combining the positive forces from the seven domains mentioned previously. Compared, e.g., with energy consumption where one number is enough, maybe placed on a graded scale for visual comparison, and measured with well known instruments, 
privacy involves multiple concepts (thus we have \ReggyName\ and \LanceyName\ with us) spanning from social to technological, most of which we do not know how to measure (here \TechyName\ could help), definitely not in a universal way since they are relative to a person's view on privacy 
(which \UpsyName\ can tell us more about)
influenced by this person's level of knowledge about privacy (which \EddyName\ is preoccupied with). 
It is generally known that some big-tech companies are monetizing on this ignorance, but not all, usually not the SMEs, many of which would like to be able to promote their privacy consciousness as a value added to their products (isn’t that right, \BussyName?).
Our \emph{PL knight} must go through a series of challenges to prove himself (these are what we mark as \textbf{G}oals)
and will have to build up a set of skills (marked as \textbf{C}haracteristics or \textbf{O}pen \textbf{P}roblems), before the seven helpers can deem PL worthy.

\longPaper{
\subsection{Evaluating the Situation of Princess Privacy}\label{sec_evaluating_situation}

\Multy%
First we want to make clear one distinction, because people often confuse privacy with security (\Techy%
programmers do this quite often). 
\Reggy%
Although multiple standards and certifications currently do not make such clear distinctions (still looking mostly at security),
security should only be a baseline, 
e.g., GDPR considers security as one of its several data protection principles (see Art.~5~I~(f)). 
Privacy protection goals include the classical security protection goals confidentiality, integrity and availability (CIA), and in addition also privacy goals such as transparency \citep{murmann2017tools}, intervenability and unlinkability that go beyond CIA \citep{hansen2015protection}.
\Upsy%
Moreover, when looking at the attitudes of the users there are clear differences between security and privacy, due to individual differences 
\citep{egelman2015predicting}.

\Bussy%
If in security it is often said that the weakest link is the user, in privacy we see that the weakest link is the controller. Examples of privacy breaches for which the controller is responsible can be: the controller ``tricks'' the users into giving more data than the user is aware of, often through hiding information or by using privacy-invasive approaches known as ``dark patterns'' 
\citep{bosch2016DarkPrivacy,DarkPrivacy2019CHI,DarkPrivacy2020CHI};
lack of legal competence when drawing contracts with third parties; programming incompetence incurring leakage of data, e.g., usage of third party libraries; or not investing in measures for preventing security leakages \citep{palombo2020AnEthnographic}.

}

\Reggy%
The Recital 100 of GDPR encourages ``the establishment of certifications mechanisms and data protection seals and marks [to allow] data subjects to quickly assess the level of data protection of relevant products and services'' \citep{EU2016GDPR}.
While Art.~42(1) encourages the implementation of certification and data protection seals for demonstrating compliance by accredited certification bodies,
PL should go beyond and measure on a scale how well the privacy is respected and how easy is for a user to understand that (see also \refGoal{goal_usableGDPR}).

\begin{goal}\label{goal_onExistCertif}
One \textbf{G}oal is to build PL on/into existing certifications.
\end{goal}

\Reggy%
Examples of existing certifications include:
Datenschutzgutesiegel, granted to systems and products by ULD (The Schleswig-Holstein Data Protection Authority)\footnote{\url{https://www.datenschutzzentrum.de/guetesiegel/}},
EuroPriSe\footnote{European Privacy Seal, \url{https://www.euprivacyseal.com/EPS-en/Home}.},
Common Criteria\footnote{\url{https://www.commoncriteriaportal.org/}} (ISO/IEC 15408) including a Privacy Class meant for defining privacy functionality, focusing on aspects such as anonymity \citep{ELLIOT2018funcAnonym}, pseudonymity, unlinkability \citep{MADAAN2018likability}, unobservability \citep{Pfitzmann2001}.
\Lancey%
Some of these are partly required by law, e.g., ULD Datenschutzgutesiegel is used for public procurement in the Schleswig-Holstein German state,\jj{ [Maybe a bit more about this from Harald?] } whereas Common Criteria are taken up in 
eIDAS (electronic IDentification, Authentication and trust Services) EU Regulation No 910/2014,
and partly required for certain procurements in certain public sectors.

\Reggy%
For privacy certifications we also need evaluations. The challenge is that the focus of schemes 
such as the above
is much on security testing and penetration testing. There is a need for more formal evaluation, verification, or testing of privacy requirements (as \TechyName\ can soon tell more about).
\Upsy%
Aspects of usability should also be included in the evaluation of privacy. Parallel this with security where the weakest link is often the end-user. Nowadays communication protocols are formally proven secure, but still security breaches occur because the user interfaces are not usable, leading end-users to doing mistakes, e.g., the security warnings for SSL certificates or other types of security warnings where the end-users have to make decisions without good guidance or usable instructions, e.g., Whitten and Tygar already in 1999 tested Pretty Good Privacy and revealed several usability issues that lead to insecure decisions or that the encryption products/features were not used at all \citep{whitten1999johnny}.%

\longPaper{
\Upsy%
In addition,
privacy (like security) is usually only a secondary task for the users \citep{whitten1999johnny}, e.g., when buying train tickets with a ticket-app the primary goal of the user is not to check how well the app protects her privacy, but to reach a certain destination. 
In addition, it is arduous for a regular person to keep track of all the electronic data that she is generating, given that many activities nowadays are happening online. It is even more difficult to know exactly which effect this data has on our privacy, because of the modern machine learning algorithms that can make inferences based of apparently non-private pieces of data \citep{rader2020NarrowThought,acquisti2017nudges}.

\Upsy%
To overcome such user/usability related challenges, one has to make the privacy related measures usable. 
\Lancey%
The GDPR is a good place for finding examples of usability goals, e.g., ``communication ... relating to processing [to be provided] to the data subject in a concise, transparent, intelligible and easily accessible form, using clear and plain language'' (Art.~12); with 30 more such usable privacy goals identified in \citep{johansen2020making}.
}
\longPaper{
However, the usability goals appearing in GDPR are too general, given the inherent nature of the GDPR (and laws in general), 
which in this case allow for too much subjective interpretation by controllers for their own interest.
One classical behavior is to make the privacy settings blend into the background, or even worse, the button for the privacy invasive option is highlighted by design; e.g., when emphasizing the `Accept' button for the privacy policy, the users will give their consent without reading.%
This is the case with the privacy policies as well, which can containing information manipulated in a way that the user on average will not be well informed, though they formally meet the requirements of the GDPR \citep{karegar2020dilemma,cranor2008CostPrivacy}.
Therefore:
}

\begin{goal}\label{goal_usableGDPR}
PL should offer a way to reach the usability goals of GDPR.
\end{goal}

\Upsy%
Usability and Human-Computer Interaction (HCI) techniques have mostly been developed for making technologies that are difficult to use, or made for highly specialized experts, more easy to understand and interact with, both for the expert user, but often also for new, less expert users. Usability is even more important for privacy since privacy is a highly complex concept, related to complex technologies such as AI and Big Data, but which is especially addressed to the laypeople, not to experts, because privacy is a human right.

\longPaper{
\Lancey%
As shown by \citep{patrick2003privacy,patrick2003human}, it can however be challenging to map legal requirements into HCI requirements. \TechyName\ agrees that for the programmers as well it can be difficult to implement the statements made in the privacy policies or regulations. Therefore, help is needed in bridging the gap between regulations or legal documents (such as privacy policies) and the software/technology that these talk about. Those needing support in this case being the lawyers, interaction designers, and the programmers.
}

\startingPoint{}
for evolving certification schemes from seals and trust marks towards privacy labels of the energy efficiency type, i.e., aiming for goal~\refGoal{goal_onExistCertif}, is by measuring the usability of privacy using HCI methods, thus covering also \refGoal{goal_usableGDPR}. 
\citep{johansen2020making} works in this direction by proposing a set of criteria thought to produce measurable evaluations of the effectiveness, efficiency, and satisfaction  with which privacy goals of GDPR are reached. 
This work extends the methodology of EuroPriSe certification scheme by adding, what is called, usable privacy criteria. Thus, the EuroPriSe certification assures that the GDPR legal ground is covered, including data protection principles and data subject rights, while the usable privacy criteria come on top, fine-graining the EuroPriSe evaluation  
with usability measurements showing how well the legislation is respected.%
\longPaper{ All these are organized and visualized as a cube, called ``Usable Privacy Cube'', composed of three variability axes containing: usable privacy criteria, rights of the data subjects, and privacy principles.}
This work has identified from the GDPR text 30 usability goals, which have been used as guidelines to define 23 usability criteria, each composed of several subcriteria designed to measure usability aspects \citep{johansen2020theTR}.

In ergonomics and HCI, the context of use is an important component of a usability evaluation.%
\longPaper{ Consider the definition from the ISO standard 9241 \citep{ISO9241-11:2018}: ``The context of use comprises a combination of users, goals, tasks, resources, and technical, physical and social, cultural and organizational environments in which the system or service is used.''} 
The context of use is translated into GDPR vocabulary as the \emph{context of processing}.%
\longPaper{Recital (71) of GDPR states ``In order to ensure fair and transparent processing [the controller should take] into account the specific circumstances and context in which the personal data are processed [...] .''. An example of a context eliciting question is ``What data is the product/system processing?'', which elicits information about type, volatility, accuracy, size/amount, persistence, value of the data.}
Creating guidelines for how to establish the context of processing is a necessary enhancement of the above work.

\begin{goal}\label{goal_context}
PL should reflect and communicate the context of processing.
\end{goal}

\longPaper{
\subsection{The need for a Privacy Labeling knight}\label{sec_need_for_PL}
}

\Multy%
Besides being adaptable to different contexts, 
the PL knight should also be 
trustworthy (hence \refGoal{goal_trustFactor}) and
economically inspiring 
(hence \refGoal{goal_people_economy}).

\Reggy%
In 2013, the European Consumer Centres' Network published a trust mark report “Can I trust the trust mark?”\footnote{\url{https://ec.europa.eu/info/sites/info/files/trust_mark_report_2013_en.pdf}} where it is brought to the attention the importance of establishing reliable trust and demanded a more uniform practice of European trust marks.

\Upsy%
The research done in projects such as PRIME \citep{camenisch2011PRIMEbook} and PrimeLife \citep{fischer2011PrimeLifeBook} has shown that the end-user is having a lack of trust especially in the case of PETs (Privacy Enhancing Technologies) \citep{alaqra2018enhancing}. They often have difficulty believing the claims made by the PETs that privacy can be really protected in that way, often because these are counterintuitive.
\longPaper{
\Techy%
PETs are not trusted or understood because they are based on cryptography and cryptographic schemes do things that are not easy to grasp. The user testing in PrimeLife\footnote{\url{http://primelife.ercim.eu/}} showed that people had difficulty understanding and believing the concept of data anonymization via zero-knowledge proofs. There are also no good real-word analogies/metaphors that can be used to mediate these functionalities, which seem to be counterintuitive for users \citep{wastlund2009pet}.
}

\begin{goal}\label{goal_trustFactor}
PL should be developed and applied uniformly so that it becomes an important trust factor. 
\end{goal}

\longPaper{
\Upsy%
Interviews with stakeholders involving privacy enhancing data analysis on encrypted data (homomorphically encrypted data) were done in the PAPAYA\footnote{PlAtform for PrivAcY preserving data Analytics \url{https://www.papaya-project.eu/}} project. The scenario was that ECG (electrocardiography) data were sent to the cloud for data analysis. The ECG signals were encrypted, while the analysis was taking place only in encrypted form.  In expert interviews, the more technically skilled users showed skepticism towards this form of analysis. 
The expert users had requirements to have assurance guarantees that data analysis on encrypted data really worked. When shown a privacy impact assessment (PIA) according to the tool from the French Data Protection Commission (CNIL) to increase trust, they also wanted to have complementary information about the PIA method and process, and qualification of the evaluators. This shows, for the case of expert users, the importance of privacy claims for establishing trust in PETS \citep{alaqra2020using}.
}

\Bussy%
Increasing customers' trust can also be achieved by showing that the organization takes privacy seriously, by displaying privacy information that can be understood, instead of a long legal text. Privacy governance can in this way become a competitive asset, an unique selling proposition \citep{hoffman2014privacy}.

\begin{goal}\label{goal_people_economy}
PL would facilitate the inclusion of people in the new data economy.
\end{goal}

\Bussy%
There are many reasons (and controversies) for including people in the new data economy \citep{jentzsch2012study,acquisti2013privacy,spiekermann2015vision,acquisti2016economics,li2017theory,benndorf2018willingness,MALGIERI2018pricingPrivacy}.
To do so, one needs, besides trust, to consider how well the consumers are informed and how aware they are of the existing options. PL would achieve this by implementing, in a more accessible manner, the transparency principle of GDPR (Art. 5 I (a), 11, 12), which requires data controllers to inform their users about, among other, what data is processed, for which purposes, the legitimate requirements of the processing, or who are the recipients.%
\longPaper{Having access to such information, the consumers can make more informed choices.}
We wish to empower people to gain insight and control to make informed choices and comparisons.
The consumer can then become part of the data economy not only as an asset, but also as a stakeholder that can influence the market.

\section{PL stakeholders and their needs}\label{sec_PL_Stakeholders}

\Multy%
Privacy labels can have different purposes (e.g., for internal use, for showing compliance or only for fulfilling transparency requirements, for marketers to use for selling effectively/targeted) and be intended for different audiences (e.g., data subjects or the controllers).
However, it is probably difficult to put everything in one label.

\begin{characteristic}\label{charact_usable}
One \textbf{C}haracteristic of PL is to be usable, for different purposes, by different 
types of stakeholders.
\end{characteristic}

As any knight who is evaluated on his traits of character, \refCharacteristic{charact_usable} is only the first of many more characteristics to be argued for in the rest of this text.

\begin{openproblem}
An \textbf{O}pen \textbf{P}roblem is how to make the concept of PL flexible enough to accommodate different purposes and audiences, and for each such PL instance how should it be designed in order to convey the intended information to the intended audience.
\end{openproblem}

The rest of this section surveys the various audiences and purposes PL may have, starting with the ``convenience users'' being our primary target, and continuing to discuss the needs and expectations of businesses, lawyers, regulators and authorities, and programmers.

\subsection{Convenience users}\label{subsec_convenienceUsers}

We define \emph{``convenience users''} as those people that nowadays trade in their privacy for convenience, most often without knowing what they are trading in. 

\Bussy%
The convenience users are much of the time running on `autopilot' when they are making judgments, e.g., when shopping online.
This happens from multiple reasons, e.g., willpower depletion \citep{baumeister11willpower}, heuristic and intuitive thinking \citep{kahneman2011thinking}, or manipulations such as priming done through media channels and advertising \citep{cialdini07influenceBook,thaler2009nudge,harris2009priming}.
Advertisers and commercial businesses use extensively behavioral psychology to influence, whereas the governments and authorities seem to assume that people are rational and as such they do not see the need to push any psychological buttons. 
\Reggy%
The rational behavior assumption seems to be the case also with the current certification schemes, which are based on a rational model. They state that privacy is inherent in the product or the service as a sort of objective measurement, and they are made on the premise that one size fits all.

\Eddy%
Knowledge is power, and our PL knight can help increase the privacy knowledge of laypeople, thus empowering them to make well informed choices in the technological world and thus actively participate in the data economy.
Convenience users would get in contact with the PL as a result of being interested in buying a digital product or using a new digital service that is collecting personal data (most of which do).

\begin{goal}
PL as an educational tool to increase privacy literacy in the general population.
\end{goal}

\begin{figure}[!b]
\centering
\includegraphics[width=5.5cm]{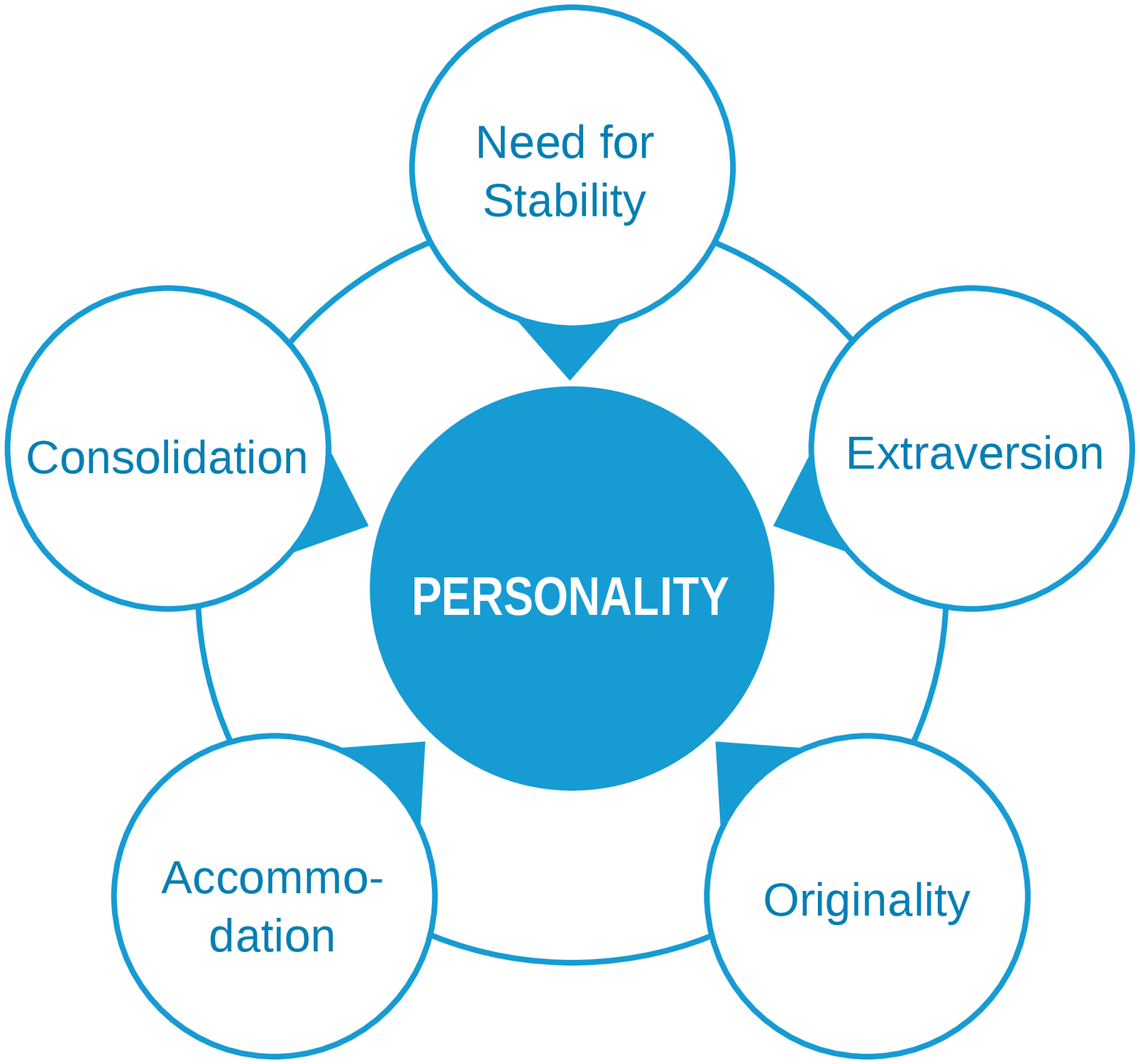}
\caption{Personality traits, from Chap.~30 of \citep{howard2014owner}.}\label{fig_personality_traits}
\end{figure}

\Eddy%
At the same time, we also know that people 
have different values and traits of personality (Fig.~\ref{fig_personality_traits}\longPaper{; see also the original Big Five model \citep{allport1937personality}}), and that different people may prefer different levels of privacy, that may also change with time and context \citep{westin1991harris,knijnenburg2013dimensionality,gerber2018explaining}. In addition, psychological studies show that people are not always able to make choices or judgments that are in their best interests, as e.g., with overly confident people that jump to conclusions without the necessary due diligence or due to anxiousness, which makes one avoid making decisions \citep{john2008paradigm,john2010handbook}. These differences in people's personality traits are susceptible to different kinds of influences \citep{acquisti2009nudging,nudge2019privacy}.

\Eddy%
People also have different cognitive styles. Some have an intuitive approach to making judgments and decisions about something, while others have a more analytical approach \citep{egelman2015predicting}.
We also know that people's judgments are very much influenced by their current emotional state, their affect, their moods \citep{kitkowska2020psychological}, 
e.g., car salespersons are known to try to get the buyer in a good mood, in order for the buyer to be less critical. Whereas when in a bad mood or on defense, one not only becomes captious, but also more analytical \citep{peters2006affect}.%
Combining the personality and cognitive style where a person is not confident and is not prone to making analytical judgments, but instead has an intuitive cognitive style and also is in a good mood, is what might be the characteristic of the typical convenience user. In this case it might be necessary to push other types of buttons to slow the cognitive processing of this type of users down, so that they can really think about what they're doing, instead of processing and making judgments intuitively and heuristically.

\Bussy%
The concept of social influence that is a part of social psychology, describes how people may be influenced by different agents, in different ways, with different means, or for different purposes \citep{cialdini07influenceBook,argo2020contemporary}. A person buying an app that was advertised on the subway, is an example of social influence that does not always serve peoples' best interests, but usually serves commercial interests. 
\Reggy%
Instead, governments or institutions (e.g., independent supervisory authorities), might be having a more ethical approach to influencing people 
to
make decisions that are in their best interests.
\longPaper{
\Eddy%
If people are not aware and they run on cognitive autopilot, then even if the person is not capable of making an analytical judgment or decision, that serves his or her best interests, they can nevertheless be pushed in the right direction, if the ones that steer have 
the respective people'
best interests in mind.}

\begin{figure}[t]
\centering
\includegraphics[width=5.3cm]{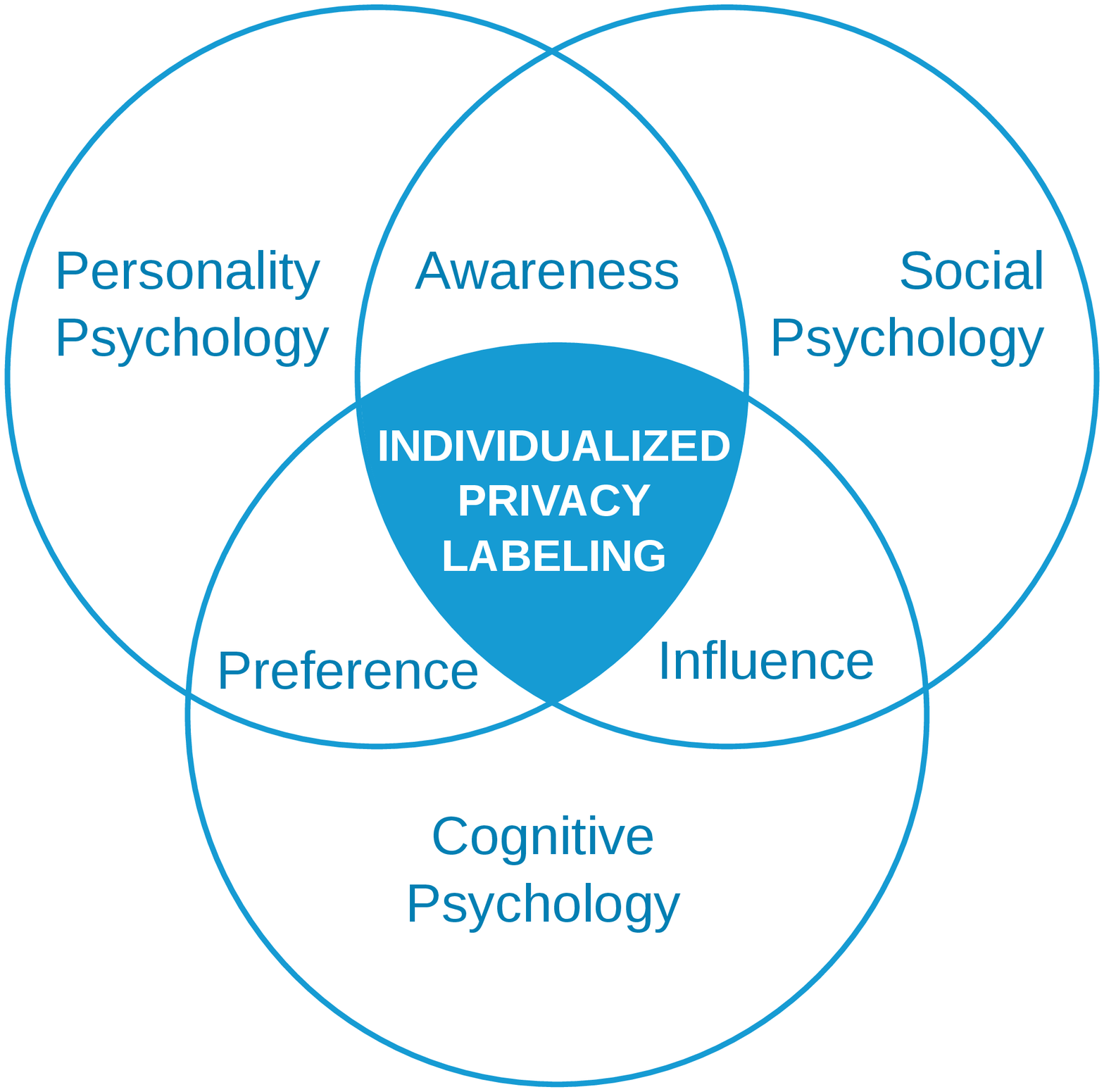}
\caption{Individualized PL.}\label{fig_individualizedPL_psychology}
\end{figure}

\begin{characteristic}
PL should be individualized by considering the psychology of personality, cognitive styles, and social influence (cf.~Fig.~\ref{fig_individualizedPL_psychology}). 
\end{characteristic}

\startingPoint{}
is the combination of the elements from Fig.~\ref{fig_individualizedPL_psychology} that would result in a more nuanced distribution of privacy preferences and attitudes. 
\Upsy%
The literature has identified several of such predominant profiles known as privacy personas\footnote{A \emph{persona} is a precise description of the user of a system and what she wishes to accomplish. Though a persona is not a real person, it is created based on synthesized characteristics and needs of real people, and it represents the profile of a typical user \citep{cooper2004inmates}.} (e.g., information controllers, security concerned, benefits seekers, crowd followers, and organizational assurance seekers) \citep{westin1967,morton2014desperately,woodruff2014would,dupree2016privacy}. Rather then adopting an exclusivist and reductionist approach, the PL should be able to adapt to different privacy preferences or privacy personas.

\subsection{Businesses}

\Bussy%
PL could provide a competitive advantage \citep{martin2017role}, especially in markets such as Europe, where privacy protections are required by law. Even in privacy unregulated markets, a company that has visible and easily understandable statements of compliance will be perceived by the customers as one that takes privacy seriously. This has the potential to increase the customer's trust in the company, which is seen in  \citep{bachlechner2020privacy} as essential for data-driven businesses.

\Bussy%
However, against the possible benefits one has to weigh the possible costs. One factor to consider is the price of obtaining and maintaining a PL, e.g., how often re-certification is required or whether the PL will reduce or increase the use of other resources, e.g., compliance officers, additional IT staff, preparing more documentation, or legal council.%
\longPaper{Another important question would ask what is the scope of the label and what does it cover, coupled with an evaluation of how time consuming is to obtain and maintain a PL or, alternatively, how much time would it save on other business aspects such as marketing or customer retention.} 
Since the benefits of a label also depend on its reputation, business are interested in, e.g., how widely known is the label, whether it is already recognizable, and if not, whether it will be so in the near future.

\begin{goal}
PL should become something worth investing in, that brings clear benefits and offers a competitive advantage, outweighing the costs.
\end{goal}

\Bussy%
Businesses tend to prioritize what is most requested by their customers. 
Recent research shows that privacy is a major blocker for the adoption of new technologies, e.g., \citep{barbosa2020soups} studied people's considerations for adoption of smart home devices, and found that half of the 613 participants have named privacy or security concerns/risks as being a blocker for them acquiring an IoT device. Privacy was also ranked second, after `convenience', as being considered when purchasing such devices.

\Lancey%
Depending on the needs of the business, e.g., in which territory they plan to distribute a new product, the privacy related documentation is adjusted considering the costs and benefits. The level of detail of a privacy policy (and of the related PL, cf.~\refCharacteristic{charact_linked_ToS} \& \refOpenProblem{open_linkToS}) depends on, e.g., what the local regulations require, but also on how complicated the processing of personal data of the respective company is, restricted by the resources available to the company (e.g., to pay lawyers or certification processes). Since privacy policies can vary widely, it is important to consider the flexibility of PL and create different modules that can adapt to more complicated needs.

\begin{characteristic}
PL must be modular and flexible to accommodate the different needs of businesses in relation to local regulations and their customers' demands.
\end{characteristic}

\Lancey%
Together with stricter regulations such as GDPR \citep{Tikkinen2018StrictGDPR} and peer pressure, 
e.g., Google Play requiring all apps to have a privacy policy before being allowed on the app store, privacy has become a more important topic for businesses.
Irrespective of whether it is a large multinational company or a start-up, every business that deals with digital data nowadays
has to have a privacy policy.

\longPaper{
\Multy%
Previously, a common practice was to just copy and paste privacy policies, e.g., a new start-up wanting to provide something similar to an existing service, say like Github, they would copy the privacy policy from Github, maybe trying to read it themselves and maybe changing a few things.\footnote{This practice has proliferated to the point that now one can find ToS generators and ToS templates as services online, see e.g.: 
\href{https://www.termsofservicegenerator.net/}{www.termsofservicegenerator.net} or \href{https://getterms.io/}{getterms.io} or \href{https://privacyterms.io/}{privacyterms.io} 
for generating various kinds of online agreements including privacy policies, 
and 
\href{https://www.termsservicetemplate.com/}{www.termsservicetemplate.com} or \href{https://www.privacypolicies.com/blog/sample-terms-service-template/}{www.privacypolicies.com/blog/sample-terms-service-template} 
for templates; 
and 
\href{https://www.termsfeed.com/blog/terms-conditions-copyright-law/}{www.termsfeed.com/blog/terms-conditions-copyright-law} 
for arguments and discussions around such services.}
\Lancey%
However, nowadays standardized privacy policies is an obsolete practice, because complying with requirements like GDPR implies providing valuable information to the user. This information cannot be general, but has to explain what the company is actually doing with personal data. The lawyers use considerable time talking with the client (who is the controller), investigating, making sure that they completely understand their operations and what is necessary, in order to make a meaningful privacy policy.}

\begin{characteristic}\label{charact_linked_ToS}
PL should be closely related to privacy policies,\footnote{One can also consider other privacy related documents or certification processes.} which the law asks all businesses to have.
\end{characteristic}

\startingPoint{}
in the process of making a meaningful privacy policy could be to employ the support from a tool, such as the NL.PL (see Section~\ref{subsec_approach_self} for more details), where one is guided into providing the information required for making a PL (and thus also a privacy policy). Similarly to what a lawyer would do, one still needs to do research ahead to extract the needed information to be used with such a tool.

\Bussy%
The NL.PL focuses on facilitating more transparency and clear communication by making the privacy statements easy to understand for customers. 
A tool like NL.PL can be used by everyone (both businesses and individuals) to have an overview of a privacy statement and be the starting point for both a privacy label and a fully compliant privacy policy. It is especially useful for SMEs, which often do not have the means to pay law firms to create the necessary legal documentation. 
The businesses can save costs by doing the needed preliminary work by themselves with the help of the tool. 
\longPaper{To help 
a privacy officer 
create a label for their own organization with ease, NL.PL follows a data flow organization model, which is provided with company specific privacy relevant details, e.g., about collected data, such as location and duration, or processing purpose (with predefined example to choose from and edit).}

\Multy%
A more general problem encompassing what has been said until now about privacy policies and businesses adopting PL is \refOpenProblem{open_massMarket}.

\begin{openproblem}\label{open_massMarket}
How can Privacy Labels reach the mass market?\onlyShortPaper{\\ \phantom{to get the icon to the right}}
\end{openproblem}

\Lancey%
It can be said that privacy policies, being required by law for every digital product or service handling private information, have already reached the mass market. By attaching PL to privacy policies, these too would reach all the consumers under the condition that businesses would be willing to adopt PL. 
\Bussy%
Therefore, in an unregulated market place, PL has to appeal also to businesses, to bring value to their products. 
\Reggy%
Otherwise, regulations and legislation, along with incentives, can be used to attain a critical mass of businesses using PL, at least as a means of making their privacy policies more usable.
\Bussy%
A tool such as NL.PL would also be useful in this respect, as it would make the creation of PL more practical and affordable, 
through 
being easy to use without needing legal expertise.

\subsection{Lawyers}

\Lancey%
A common situation relevant for lawyers, is when a client is considering obtaining a privacy label.
The primary choice is a privacy label that would also involve a certification by an independent certification body, e.g., if the client is established in the EEA (European Economic Area) then it will have to adhere to data protection legislation such as the GDPR and also local data protection requirements. However, PL could also be relevant for establishments outside the EEA, for example if they want to compete with businesses in the EEA.
Since there are not many privacy certification options, it is also often that PL are desired rather as information conveying tools. 
Since privacy legislation (the same as certifications) can vary between different geographical regions \citep{SULLIVAN2019GDPRvsAPEC,Kaminski2020CACM_privacyLaw}, one can see PL as a harmonizing factor because of its international nature, managed by a global community (as we detail in Section~\ref{subsec_approach_crowd}).
The type of client is important as well, with factors such as the business size, multinational, national, or an SME, combined with the nature of their commercial transactions. A PL can thus be relevant both for B2C (Business to Consumers) as well as B2B (Business to Business). 

\begin{characteristic}
PL must take into consideration both commercial and legal aspects, and their interdependencies.
\end{characteristic}

\Lancey%
In terms of legal implications, having a PL does not relieve the company of its obligations to adhere to the data protection law. PL is only a modality to demonstrate compliance, as stipulated by GDPR in Art.~42(1). 
Therefore, besides maintaining documentation relevant for PL, the company still needs to implement technical and organizational measures, which should not be seen as something additional to having a PL, but as part of the requirements necessary for obtaining the PL.

\begin{characteristic}
PL should reflect technical and organizational measures taken by the company.
\end{characteristic}

\Bussy%
Certifications sometimes have additional requirements that are more onerous than the law. It can be more difficult to obtain the PL through a certification process than to just be compliant with data protection law. This could have potential additional costs but also potential benefits. 
Having a PL might have implications for other business areas where legal advise is usually needed, as in marketing where the lawyer has to assist in ensuring that the PL is not misleading or inaccurate, otherwise it can be judged as false marketing.

\begin{characteristic}
PL should come with supporting guidelines for businesses so that they do not include false information unwittingly.
\end{characteristic}

Too often in fairytales, malicious characters take the form of, or pretend to be, the good characters, like Snow White's stepmother who disguises herself as an old peddler or a comb seller in her attempts to kill Snow White. 

\begin{openproblem}
We should develop a system to distinguish between a false and an authentic PL. 
\end{openproblem}

\longPaper{
\Reggy%
Standardization is needed to make services comparable. 
We wish for a basic way to structure the privacy related aspects in a fixed and similar manner. Such a structure allows also for cross-comparison of labeled products and services. 
\Lancey%
Having an overview is always useful for a lawyer as well as for convenience users. 
It can be sometimes discrepancy
in the needs a lawyer might have, e.g., the requirement of being transparent is not always appreciated for some lawyers, because they might want to have room to maneuver in case something goes wrong. However, PL aims to be a standardized and clear approach to presenting information towards customers and consumers. A lawyer or a consultant representing a company is required in this way to simply be transparent and demonstrate compliance.
}

\subsection{Programmers}

\Lancey%
Technology people are struggling to understand the legal terminology and how to implement a system so to conform with the statements appearing in the legislation and in the privacy policies made by their leadership.
This ``legal-text-to-code'' gap is even larger than the well-known gap between software requirements (or specifications) and their implementation.
\Techy%
Besides standard questions that programmers ask, such as ``What does data minimization mean?'', 
one important problem that they face is how to match the ``purpose'' stated in a privacy policy (and presented by the PL) with the precise usage of the data during any execution of their software implementation.
These are necessary questions when trying to enforce privacy or prove compliance with the GDPR or (maybe easier) to own privacy policies. It is already difficult for lawyers at an organizational level to deal with such questions, which become even more complicated to answer when trying to look at the software code.

\begin{openproblem}
If PL use tools to translate/explain privacy policies to convenience users, we would like to investigate how can these same tools be useful to the programmers to understand how to implement the statements from privacy policies and law.
\end{openproblem}

\startingPoint{}
can be to try to use existing formal tools to analyze at least the more critical parts of the code by, e.g., doing code inspection.
Since it is difficult to analyze code automatically, putting a human expert into the loop can be a more feasible first approach of doing semi-automated code evaluation  and verification for privacy compliance.
\Techy%
There exist several recent technological advances on  automating particular aspects of
privacy, e.g., on data-flow \citep{antignac2016privacy}; on data minimization \citep{antignac2017data}; on privacy by design \citep{langheinrich2001privacy,gurses2011engineering,hoepman2014privacy,ROMANOU201899,antignac2018privacy,schneider2018privacy}; and in general on Privacy-Enhancing Technologies (PETs) \citep{DanezisDHHMTS15}.
However, it can take long for a research idea to reach the programmers, and even more so for PETs since more often than not, these prove too difficult for software development companies to comprehend, let alone implement or adopt in their software or DevOps tool-chains.

One of the fastest spreading types of software is the AI/Deep learning based software (see the 2019 Turing award laureates excellent overview \citep{AIinNature2015}), with a considerable number of programmers actively involved. In his recent call for AI regulations, Etzioni urges regulators to focus on five critical areas ``no killing, responsibility, transparency, privacy, and bias'' \citep{Etzioni2018CACM}. Privacy has earned a forth place on this list of concerns for AI software because AI is data-hungry and much of this data will presumably come from IoT systems close to humans -- of course, for those AI applications that are interacting in some form with people and the society at large (e.g., in decision support or smart-* systems). 
Privacy labels could help in these regulating endeavors as well, this time not addressed only to the convenience users but more to the businesses, e.g, for allowing well informed AI software purchases, as well as to AI engineers, e.g., to guide their choices of libraries and software components.

\longPaper{
Privacy has long been an important concern for software developers, e.g., the (then) President of the ACM, David Parrerson in \citep{patterson200520th} put forward the ``SPUR manifesto'' which was placing (P)rivacy as one of the four main focus areas for software engineering, along with (S)ecurity, (U)sability, and (R)eliability. On the contrary, biases in software (chiefly in AI-based decision systems, as Etzioni describes) is a rather new concept for software developers.
}

Biases in AI systems (the fifth area of concern for Etzioni) normally come from improper use of training data, i.e., the system is trained on a data set that is not representative of the population/problem that it is applied to (or does predictions about). AI biases can be about gender, race, or other social aspects \citep{caliskan2017semantics,zou2018AI,silva2019algorithms}, but also about privacy. This last form of bias, which we call \textit{privacy biases}, is largely not investigated because it is not seen as a machine bias, i.e., it does not appear from data or the software code. Privacy biases are human biases, in line with the traditional bias mechanisms studied in psychology \citep{gilovich2002heuristics,tversky1974judgment,oliver2014satisfaction,wilson2003affective} -- see also a nice account of how cognitive and behavioral biasing mechanisms (such as the anchoring heuristic or framing effect) influence privacy behaviors in \cite[Sec.2.3]{acquisti2017nudges}.
Quite a number of privacy biases could fall in the class that we would call \textit{I-have-nothing-to-hide}, with a large collection of such privacy attitudes nicely presented in \citep{solove2011nothing}.
A privacy bias that programmers often fall pray to can be called \textit{privacy=security}, which we have already explained in the begging of Section~\ref{sec_evaluating_situation}.
Recent results \citep{pedersen2020studying} have shown that human biases can be \textit{transferred} from the programmer into the software that she is building. 

Bias transference is thus an additional mechanism to the standard one studied in AI biases, through which human biases can manifest into the software that we build. Therefore, privacy biases are elevated to being a serious threat to the software that programmers develop as it can incorporate the privacy biases of their creators.
Privacy biases have as a root cause the lack of adequate knowledge, either that the person is time constrained and cannot gather or infer the needed knowledge for the decision task at hand, or that simply the person is inexperienced for the new task.
Programmers often find themselves in such uncertainty situations, e.g., when faced with incomplete specifications or vague requirements. This is even more so in the case of understanding privacy policies.
As a result, programmers are mostly left to their own means and judgment when implementing privacy features or requirements; and any privacy bias or neglect can reflect on the users of the resulting software. This happens because of the transfer of the programmers privacy views and biases into the software artifact \citep{pedersen2020studying} when privacy aspects are not easy to comprehend.

\begin{openproblem}
How can PL prevent the transfer of the programmers' privacy views and biases into a source of privacy problems in software?
\end{openproblem}

Since PL would be associated to privacy agreements (cf.~\ref{charact_linked_ToS}) and having one goal to explaining concepts such as purpose of processing (see Section~\ref{sec_PL_for_Edu}), they would help programmers to better understand those aspects from the privacy agreement that are relevant for the product they are building.

\subsection{Regulators, Certification bodies, and Authorities}

\Reggy%
Certifications bodies as stakeholders can see the PL as a means to convey their certification results. The provisions in Art.~42/43 of GDPR strengthen the role of the certification bodies as a means for the companies to show compliance \citep{LACHAUD2018certification}. 
PL should contribute to the further development and enhancement of the existing certification schemes (cf.~\refGoal{goal_onExistCertif}).

Data Protection Authorities (DPAs) tend to rely on detailed sources, such as privacy policies and technical documentation, in their audit work. Therefore, highly relevant for DPAs would be the deeper layers of the PL, where detailed information is offered (cf.~Section~\ref{subsec_layered_PL}). 
However, DPAs are also responsible with checking if the visual and the ``surface'' components of the PL are an accurate reflection of the privacy policies and actual practices. In this case, their auditing work could be simplified through the automation tools and process used to 
generate the PL (cf.~Section~\ref{sec_automation}).
\longPaper{Furthermore, their work becomes universally valid if the same tools, practices and methods are used across all services. Uniform practices is one of the goals (\ref{goal_trustFactor}) we set in this paper for PL.}

DPAs are also part of the Data Protection Board where they can interact with privacy regulators on various aspects of the legislation and its applications. Privacy is a concern in various social/economical areas, such as health, with a major role to play in the future of AI regulations \citep{Etzioni2018CACM,CLARKE2019AIregulations}. PL could be introduced as an essential aspect of such regulations since PL allows easy comparisons regarding privacy between AI systems.

Data-intensive technologies such as AI-based decision systems or management software have entered also in the many state institutions such as in policing \citep{brayne2017big} or courts \citep{cscw2019Recidivism,dressel2018accuracy,MALGIERI2019105327}. Privacy is maybe of a greater concern to such institutions than it is to companies. In state institutions, decisions on purchasing a piece of technology or service is done through a highly regulated and transparent process called procurements. Privacy would thus be part of the requirements mentioned in the procurement call. PL could also here be used to make it easier to evaluated the proposals. This is the same way of applying PL in any form of technology purchase decision, be that done by a convenience user when looking to by a new IoT device, or a company management person looking to acquire a new service, or a governmental institution in a procurement process.

\begin{goal}\label{goal_procurement}
In public/private procurement, PL could be an advantage or sometimes even a requirement.
\end{goal}

Data Protection Officers (DPOs) are, according to Art.~37-39 of GDPR, acting as intermediaries between the supervisory authorities, data subjects, and the organization by which they have been appointed. The role of DPOs is to facilitate compliance, and, besides certifications, are another instrument that can be adopted by companies to ensure accountability. In some cases, GDPR makes appointing a DPO mandatory, while for the rest of the organizations this is voluntary, in which case the organization may choose to use external DPOs. This is already the case in countries such as Germany, where there is a large community of external data protection officials hired by companies. However, there are differences in the level of use of the external DPOs between the countries. In the Netherlands, for example, the companies chose to handle the compliance mostly by themselves, using external experts only for one or two days per month. 
DPOs as stakeholders for PL would have commercial interests to foster self-evaluations (expanded upon in Section~\ref{subsec_approach_self}), where they could provide companies with input.

\section{PL as a means of education and behavior change}\label{sec_PL_for_Edu}

The \textit{New Chicago School} model \citep{lessig1998new} explains how there are several \textit{modalities of regulation} for the behaviors of people, and we would also argue that it applies to businesses as well. PL are meant to help regulate the behavior of convenience users when making choices that might influence their privacy, as well as regulating the behavior of businesses that handle private data. Therefore, the multidisciplinary  character of PL involves multiple stakeholders, besides the law and regulatory institutions, in driving privacy behavior changes.

\begin{openproblem}
One open problem that PL could be useful for is to help change the behaviors and attitudes of people in regard to privacy.
\end{openproblem}

\Eddy%
The Prochaska model of stages of behavioral change \citep{prochaska1997transtheoretical}%
, often used to change behavior of addicted people, identifies several stages of awareness and appropriate actions. 
One may not be aware at all that she needs to make behavioral changes, meaning that the action targeting this person is to raise her awareness, e.g., in regard to privacy aspects.
Then the person moves into the contemplation phase when realizing that there is an important concern, e.g., privacy, which she needs to think about. Having learned and understood the problem, the person has to determine whether, and what, to do. This is a decision point where PL can help, e.g., when the person needs to take action when buying a digital product.

\Multy%
Many people are not even aware of their lack of privacy. 
We see the PL as a tool for raising awareness. 
We already have examples from the food industry where labels are used to raise awareness about the quality of the food.%
\longPaper{\footnote{``The Keyhole for healthier food'' is a Nordic voluntary label for food, introduced in 2009 as a device for raising awareness in the population towards making healthier food choices.  \url{https://helsenorge.no/other-languages/english/keyhole-healthy-food}}
For example, people may be accustomed to think that all foods are healthy, but by seeing the labels they realize that some foods are healthier than other.
They might try to find out more about the meaning of the label and might start discussing it. 
Awareness towards specific characteristics of digital products are similarly triggered by displaying different labels.
Privacy labels, when attached to, e.g., mobile apps, and are visible in an app-store or comparative table, they could be the starting point for people to realize that one product is different from another when it comes to privacy protection. Only then people might go and look for further information about the meaning of the label and its contents.} 
However, such markings can easily be used misleadingly for commercial purposes as well. 
\longPaper{If it is not a standardized label with clearly established frames, but one that the businesses choose to give the product by themselves, the package of the product might emphasize some aspects and omit others, e.g., displaying that the food contains 30\% less fat, but not saying that it contains 30\% more sugar.} 
Therefore, if not regulated, PL could be used in a suggestive and deceptive manner to induce subjective and irrational (i.e., inappropriate) behavior that is not in the user's best interest.

\Reggy%
A simple seal/label can be used to raise awareness. However, we would go beyond a mere seal. We envision a privacy label that displays information through witch people can learn more about privacy related aspects, e.g., that location sharing is a privacy sensitive information, or information about 
\longPaper{how much data the provider is collecting from their subjects and} 
what kind of data is being collected and processed, and for what purposes. The label can thus be the point of entry, providing the information that can be used to further educate people.

\begin{goal}
PL aims first at raising awareness and then further increasing knowledge and understanding of privacy in the population.
\end{goal}

\Eddy%
According to the theory of planned behavior \citep{ajzen1991theory}, behavioral change and judgments that one makes are conscious and are planned. This is a rational approach to behavioral change and decision making. However, it is known that people do not always make decisions rationally, 
e.g., when in time constraint or when one has insufficient or too complex information, one will not be able to carry out this type of rational approach to solve a new problem \citep{tversky1974judgment,kahneman1991anomalies,acquisti2017nudges}.
An alternative model is that of nudging \citep{thaler2009nudge}, which is an empirical approach to behavioral change
exploiting the automatic, heuristic-based, intuitive thinking of \citep{kahneman2011thinking,gilovich2002heuristics}. 

\Reggy%
Even if nudging for privacy is debatable because, e.g., it may restrict the individual's autonomy \citep{RENAUD201822,jarovsky2018improving} as it uses psychological mechanisms covertly, functioning unconsciously, nudging may be considered ethical as long as it is used in people's best interest; presuming one knows what is actually in peoples' best interest \citep{hausman2010debate}.
\Lancey%
Having an ethical approach to nudging, and not use it for commercial purposes \citep{sunstein2017nudges,Thaler2018sludge,caraban202023waysNudgeFramework,narayanan2020dark}, one can build a choice architecture that leads people to doing the right things and to carrying out the right activities that lead to the right decisions even if they are not aware of what they are doing.\longPaper{ We already have many examples of ethical nudges used in the traffic for the protection of the drivers and pedestrians, such as speed limit signs and speed bumps.} 
\Eddy%
However, there are many recent examples of consent forms that are GDPR compliant, but still nudge users to pick the privacy-intrusive choices, see e.g., cookie-banners \citep{machuletz2020multiple,cookies2020,cookies2019}\longPaper{, or emphasized buttons that nudge the users to select all cookies, while the possibility to not `Select all and continue' have very little visibility and are ambiguous about which purpose they serve}.
\longPaper{
One other famous and old example not related to consent forms is the ``opt-in/opt-out'' check-boxes \citep{bellman2001Darksite}, e.g., for receiving newsletters or offers from the respective company after performing an online transaction such as registering for a service or shopping online. 
\Bussy%
In many cases, and especially for privacy, how the company sets the `default' checked/unchecked is done to serve the interest of the company. Such practices are known as ``dark patterns'', and often applied disrespecting privacy 
\citep{bosch2016DarkPrivacy,DarkPrivacy2019CHI,DarkPrivacy2020CHI}; 
even though the same nudging  can be used also for good purposes, e.g., to increase the number of organ donors 
\citep{Johnson2003ScienceNudgeOrgans}.
}
\Lancey%
It is difficult to control by law or regulation the use of dark patterns \citep{waldman2020cognitive,narayanan2020dark}.

\begin{openproblem}
How can PL be a privacy nudge instrument to use for helping people make more privacy-conscious decisions when choosing a product? 
\end{openproblem}

\longPaper{
One needs to distinguish between a rational approach to influencing people and an empirical nudging approach. By considering the combination of personality psychology, social psychology and cognitive psychology, from Figure~\ref{fig_individualizedPL_psychology}, one can influence people and raise their awareness towards preferences that are good for them, or one could simply nudge them into doing that, with or without them being aware of what they are doing. Nudging does not always need to be covertly. \citep{caraban2019Nudge} shows that 78\% of the nudges presented in the HCI literature make their intentions and means transparent to the user, prompting them to make an reflective choice. 

Nudging should be used predominantly for cases where it is known that people run on autopilot, and they need help with making the right decisions.
How much should a person be autonomous, and how much should she be nudged into a direction, is a question of ethical considerations. However, when running on autopilot, it should not be expected that people would make rational judgments -- indeed, privacy decisions are often not done rationally -- and thus PL nudging should act in their best interest. 
}

\startingPoint{}
can be found among the existing works on using nudging for privacy purposes \citep{nudges2014CHI,nudges2016CSCW,acquisti2017nudges}.
We need then to know who we are dealing with by considering the users' specific cognitive and behavioral characteristics. As such, empirical data needs to be collected on the prevalence of cognitive styles in different situations and about dominating personality styles from different cultures and different regions, as well as gender, age, education; all of which are known to influence users' experiences and shape attitudes related to privacy concerns \citep{Kitkowska1210990,jarovsky2018improving,Kitkowska2020soups}.
\longPaper{In one study \citep{MurmannRF19noticesPersona} from the Privacy\&Us project\footnote{Marie Sk\l{}odowska-Curie Innovative Training Network Privacy\&Us: \href{https://privacyus.eu}{privacyus.eu}}, done on users of mobile health services it is shown that the notification preferences of these users correlate with their privacy personas.  Another study shows why the privacy of certain groups should be considered differently to those of the wider community.}
A test instrument could be created, where people are asked to answer a few questions, that will help with placing them in one of these domains. Furthermore, such instruments could also help raise the awareness of the users about who they are and what cognitive style they have.
\longPaper{If we are able to raise people's awareness about privacy issues, we may even be able to steer people `away' from the `maladaptive' use of mental heuristics in situations where heuristic thinking is less appropriate, and instead steer them into a more rational way of thinking, maybe even approximating a more consciously planned behavior \citep{ajzen1991theory}.}
Consider, e.g., how PL could appear different to people with high curiosity personalities compared to someone that travels much and might be interested only in location aspects, e.g., whether location is shared\longPaper{ and with whom}. 

\begin{openproblem}
We need to understand how to make the same PL slightly different to best match the needs of different types of personalities or activities.
\end{openproblem}

\vspace{-3ex}
\section{Automation and tools for creating PL}\label{sec_automation}
\label{section:AutomationAndTools}

\Lancey%
One way for having PL legally binding is to tie them to privacy policies (previously included in the Terms of Services, or ToS\longPaper{\footnote{See a community effort on explaining ToS at \url{https://tosdr.org/}}}). 
\Techy%
Machine learning and formal reasoning methods can be used to build tools to help translate (more or less) automatically between ToS and PL.

\begin{openproblem}\label{open_linkToS}
How can Privacy Labels and privacy policies be correlated?
\end{openproblem}

\Lancey%
The adoption and use of such tools in law firms depends on how inclined these are towards new technologies. Law firms foremost have the client's best interest in mind, and any tools that they adopt should serve that purpose.

\begin{figure*}[tb]
\begin{center}
\includegraphics[width=15cm]{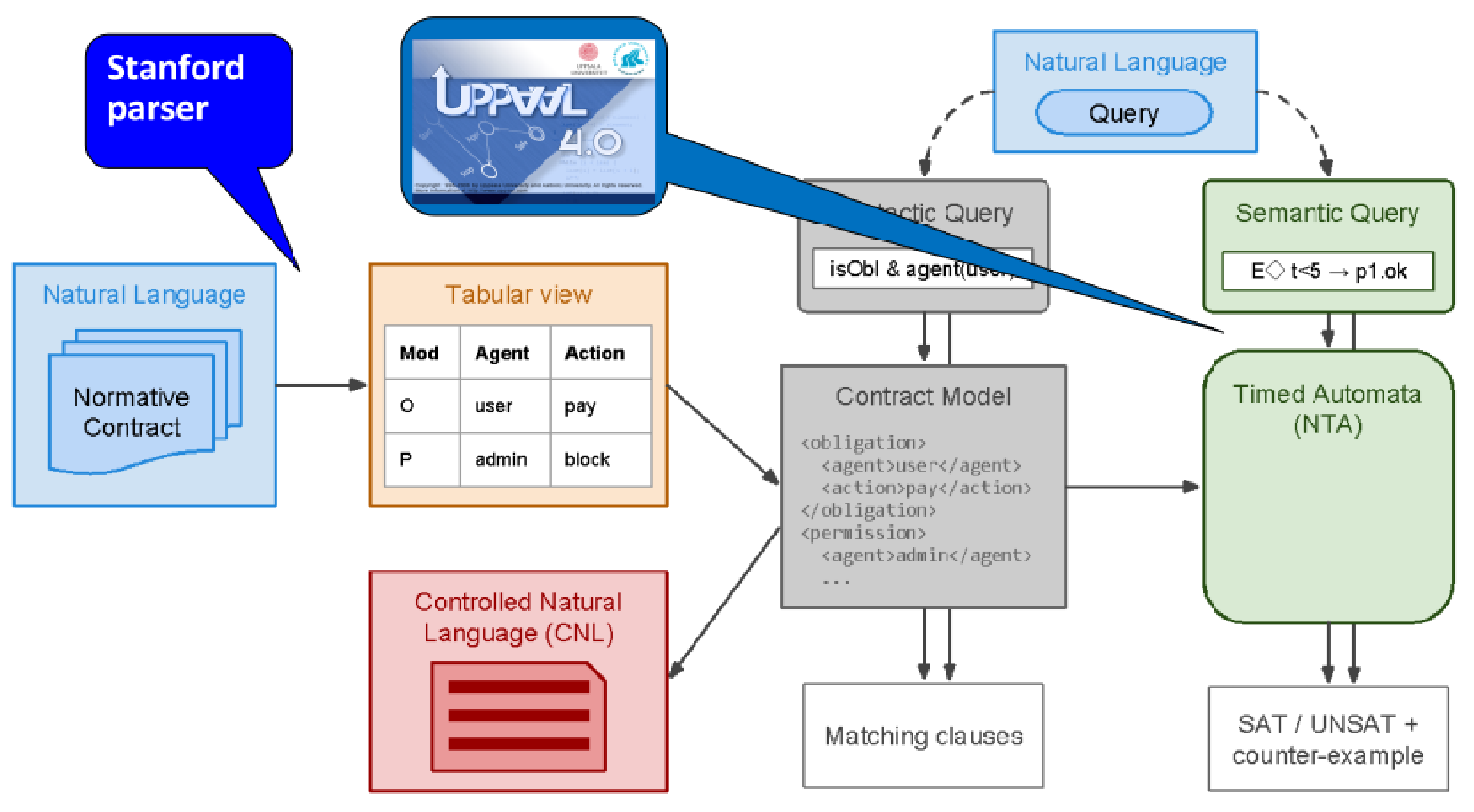}
\end{center}\label{fig_contract_Verifier_architecture}\vspace{-2ex}
\longPaper{\caption{Contract Verifier architecture for analyzing normative documents \citep{camilleri2017modelling}.}}\onlyShortPaper{\caption{Contract Verifier architecture \citep{camilleri2017modelling}.}}
\end{figure*} 

\startingPoint{} 
for research and tools relevant for translating between privacy policies and privacy labels can be found in the area of logical/legal reasoning 
and controlled natural languages. 
\Techy%
One of the more advanced tools is the Contract Verifier\footnote{\longPaper{Proof-of-concept: }\url{http://remu.grammaticalframework.org/contracts/verifier/}} \citep{camilleri2017modelling,camilleri2018web},
using several different technologies and off-the-shelve tools (see architecture in Fig.~\ref{fig_contract_Verifier_architecture}), and taking input a
\longPaper{contract, or any kind of }%
normative document written in English. Using a standard natural language parser for English (e.g., the Stanford parser\longPaper{\footnote{\url{https://nlp.stanford.edu/software/lex-parser.shtml}}}) it generates a tabular view of the different clauses in this contract. 
\longPaper{ This tabular view can be edited manually because the parser sometimes cannot parse the whole text or identify all the aspects, since the parsing from natural language into a formal model is an undecidable property (i.e., it is impossible, in general, to write a program that can do this translation fully automatically). }%
After having a tabular interpretation, everything is automatic. A formal model can be extracted and used to do queries, e.g.: what are all the obligations in the contract%
\longPaper{, or privacy policy, for a given party}; whether there are obligations without deadlines;
what is the data processing purpose; or which party the data is being share with.

\longPaper{
The Controlled Natural Language (CNL), placed in the red box in Fig.\ref{fig_contract_Verifier_architecture}, is the part where the privacy policy's natural language text is simplified into a CNL version that still looks like natural language, but is more structured and with a limited vocabulary. CNL is still readable, while also amenable to formal manipulation, which is essential for doing automated reasoning.
One problem encountered when trying to formalize legal documents is that it is very difficult to do anything unless you have a very precise definition. When one looks into the normative text -- any privacy policy written in natural language or the GDPR for the same matter -- many questions arise because words such as `adequate' or `efficient' are used, and do not have a precise meaning for technology people. When wanting to do something technically, one needs to have precise meanings for these kinds of words. One approach can be to take the usability related definition of such words, where the Usable Privacy Cube of \citep{johansen2020making,johansen2020theTR} identifies measurable criteria for such concepts related to privacy.}

\longPaper{
\Multy
Such works on automatically generating PL combined with the psychological model from Fig.\ref{fig_individualizedPL_psychology}\longPaper{ in Section~\ref{fig_individualizedPL_psychology}} would fit well within the general agenda of Behavioral Computer Science \citep{pedersen2018behavioural} since for PL both models of human behavior and of computers are needed in combination.
}

\label{subsec_AI_automation}

\Techy
Constructing PL from ToS using AI is a two-step process: 
(i) use natural language processing (NLP) to identify privacy concepts in the ToS text and 
(ii) apply a set of rules to the identified concepts to generate the relevant labels. 

\startingPoint{}
is the award-winning tools from \citep{harkous2018polisis}.
One of the few datasets that we consider to be useful for such purposes is the OPP-115 Corpus \citep{wilson2016creation}, which contains 115 privacy policies that were manually annotated by law students and then checked for inconsistencies. Each privacy policy is segmented 
and each text segment is annotated with privacy concepts. 

\longPaper{
The privacy concepts are split into a number of categories. Each text segment fits into one or more categories, each category having a
fixed set of attributes and each attribute a fixed set of possible values. After determining that a text segment fits into a certain category, we know which attributes apply and we can determine which value each attribute has.

For example, the category “Third Party Sharing/Collection” has the attribute “Identifiability” which may have
the values “identifiable”, “aggregated”, “anonymized”, etc. This is a two-stage machine learning problem: (i) first identify which category the text segment fits in, and then (ii) for each attribute determine which of the possible values are applicable. 
This approach is used by \citep{harkous2018polisis}.

Using NLP to extract data from privacy policies poses a number of challenges. One of the challenges is getting a data set that is big enough, as labeling privacy policies is challenging in itself and requires considerable time and effort. Hence the scarcity of data sets that annotate privacy concepts. Another challenge comes from the ambiguity of natural languages, which allow for sentences that cannot be automatically uniquely interpreted. 
Some ambiguities are even seen by some to be advantageous in legal language. 
In large legal text, like privacy policies or regulations, challenges also come from is the complex document structure and potential for very long relations between sections of the text, e.g., one paragraph may state some processing purpose for the data, but later in the text there may be a list of exceptions to this statement.
}

\Lancey
Having PL correlated to ToS using NLP can be used, e.g., to compare a company's PL to their ToS to see whether they match or find discrepancies between the two. One could also convert the extracted privacy concepts to simplified natural language, thus making summaries of ToS.

The use of a set of rules that correlate between the privacy concepts annotations from the text and the elements of the constructed PL is new to standard machine learning models and can be used to explain the reasoning behind the PL's creation, i.e., if we want to use AI for our PL, we want explainable AI \citep{samek2019explainable,hoffman2018metrics,hagras2018toward}.
\longPaper{A potential use case would be a system that sorts the labels from most negative to most positive. A company can then go through the labels and decide which they would like to improve, click on the label to see the rules and the sections of the privacy policy that contributed to it, and with some domain knowledge decide what actions to take to improve the situation.}

\section{The Looks and Appearance of PL}\label{sec_looks_and_appearance}
Even if in all the fairytales the good knight has pleasing looks,
for us the appearance of PL is all about conveying information to the target group.
\Upsy%
We have much to learn from the fields of Information Design and Visualization \citep{tufte2001visual,mollerup2015data,ware2021information,few2009now,cairo2012functional,knaflic2015storytelling}, but we need to expand beyond the content being presented, to include psychology so to reach the individualized PL from Fig.~\ref{fig_individualizedPL_psychology} and to make the looks of PL useful for the educational purposes mentioned in Section~\ref{sec_PL_for_Edu}.

\Multy%
Privacy icons are important for conveying information on a first level of detail \citep{holtz2011towards,efroni2019privacy}.
Icons would be needed for each of the different privacy concepts included in the PL \citep{motti2016towards}.
Privacy icons are useful for a ``nutrition facts'' style of PL, e.g., \citep{kelley2009nutritionLabel,emami2020ask}, this approach being taken by all the privacy labels discussed below.
However, a more important part of PL would be a \emph{comparable view}, in the style of energy consummation labels, involving graded scales, with the ``nutrition facts'' design and privacy icons appearing only beneath this.

\begin{openproblem}
Evaluating the degree of data protection is complicated, but needs to be made measurable, at least for some of the privacy aspects, and fit into a graded-scale system.\onlyShortPaper{\\ \phantom{to get the icon to the right}}
\end{openproblem}

\longPaper{
It is also good to consider the recent online privacy labeling projects that have been started by different organizations or individuals.\footnote{
For the past few years the idea of Privacy Labels has caught also in the news circles, see e.g., the following opinion articles:\\
\url{https://ksr.hkspublications.org/2017/07/10/mandatory-digital-privacy-labels-one-way-to-protect-consumer-data/} 
or \\
\url{https://www.politico.com/agenda/story/2018/04/25/internet-privacy-label-000656/}.}
A few examples include:
\begin{itemize}
\item the Privacy Label that won the Gold Jury Prize in the European Design Awards\\
\url{https://europeandesign.org/submissions/privacy-label/}

\item the IoT Security and Privacy Label developed recently at CMU \\
\url{https://iotsecurityprivacy.org/labels}

\item the Privacy Nutrition Labels patterns from \\
\url{https://privacypatterns.org/patterns/Privacy-Labels}

\end{itemize}
}

\subsection{The layered information of PL}\label{subsec_layered_PL}

\Reggy
We expand on the concept of ``layered notices'' as promoted by the 
\citep{art29_2018}\longPaper{\footnote{See also the older Art.~29 Data Protection Working Party (2014), \textit{Opinion 10/2004 on More Harmonized Information Provisions}, WP 100, Brussels, 25th November 2004.}}. A layer should offer the data subject only the information needed to make the right decision at a certain moment and for a specific purpose. However, the layers in their cumulative totality should meet the requirements for compliance.
The top layer of PL 
needs to convey prominently core policy information, especially the data processing purposes, who is the data controller and other core information that has to be made transparent according to the GDPR (Art.~13). \longPaper{Moreover, information on how far security and PETs are used for implementing privacy by design should be of interest and communicated.}
\longPaper{In addition, the top layer could also contain information on the processing that has the most impact on the data subjects and that enables them to understand for each specific processing purpose what consequences it would have for them, along with any other information that could `surprise them'.} 

PL would go further and include also graded scales, e.g., inspired by energy consumption labels.
Our envisaged scenario is the following. First the user is presented with a privacy grade on a scale from A to F. The user can then click on the grade and be shown further minimum of information (following Art.~29 Working Party) and icons to explain the most important parts of the agreement, maybe color-graded to explain how these contributed\longPaper{, positively or negatively,} to the overall privacy grade. Further, the user can click on each icon to see a simplified natural language explanation of what the icon means. Finally, the user can click on the simplified natural language to be referred to the section(s) in the privacy policy the statement was constructed from. 
\longPaper{The idea is not that the user should make a decision based on a simple grade, but rather that the user can select which products they are most interested in based on the grade, compare the more detailed information and use it to decide which product they want.}

\begin{openproblem}
How can a complex privacy policy, addressing different dimensions (including purposes, data controller, data types, retention periods, etc.) be mapped into a hierarchical A-F scale, while remaining relevant for a particular context and individual user?
\end{openproblem}

Even though it may be easier to grasp, a simple letter can be misleading if it is not given in a context (e.g., the type of app, the type of application domain, the role the user takes wrt.\ the  application being evaluated). Moreover, the aggregation of the evaluations of the different aspects of a privacy policy is not easy since a policy may be more privacy-friendly in some aspects but not in others/all. Even more importantly, we want individualized privacy labels (as explained in Section~\ref{subsec_convenienceUsers} and pictured in Figure~\ref{fig_individualizedPL_psychology}) because people have different privacy preferences (reflected in their privacy personas). This implies that a privacy label grade B for Alice may be perceived like an F for Bob, and hence the PL has to dynamically change based on the privacy person it is being coupled with (i.e., presented to). 
One set of measuring scales can come from criteria regarding the usability of privacy as defined in \citep{johansen2020making}.

\longPaper{
With such a basic overview of their privacy policy the company can gain more transparency. Such a PL overview is more comprehensible for the consumer and facilitates an easier comparison of the way service providers process personal data.}

\begin{goal}
The goal with ``layered PL'' is to give the user a bird's-eye view about what privacy aspects are included in the privacy policy. Then if the user wants more information, she can drill down to a deeper level. At the bottom layer, one can find the whole privacy policy.
\end{goal}

\Reggy
Compliance seals \longPaper{such as the ones from ULD, EuroPriSe, or TrustArk\footnote{Known from 1997 to 2017 at TRUSTe: \url{https://trustarc.com/blog/2017/06/06/truste-transforms-to-trustarc/}},} usually convey only the information about the issuer of the privacy seal (and the validity period). In addition there is a document online giving full details, e.g., describing the target of the evaluation and what has been evaluated. However, such a two-levels approach (minimal seal and full detailed document) does not fulfill our desires.

Another proposal of privacy labels focuses on conveying in a concise and precise way on a single label, privacy-relevant information specifically chosen for some target group
\citep{kelley2009nutritionLabel,railean2018let,emami2020ask}.
Some elements can be clicked, to find the full details of the decision behind them.
However, these only cover parts of a complete compliance document, aiming to simplify it into a small and easily understandable label. This may work for simple products or services but it would not scale up to complicated systems or ToS.
One other source of inspiration can be the layered approach used in cookies notices \citep{cookies2019,cookies2020}, which people might already be accustomed to.

\longPaper{
Similarly, in the case of the NL.PL the layered approach chooses a number of basic elements used to produce a visual separation of the data flow -- this describes what are the sources of the data, how and for what purpose are they used in the processing activities, how long data are retained and when they are deleted. Other privacy aspects such as retention terms or security measures are also included (see Figure~\ref{fig_NL.PL_elements} for an overview).
NL.PL allows to drill down towards more details, and is built in a structured and standardized way, so that the businesses have the same topics to cover when providing privacy-relevant information during the NL.PL process.

\begin{figure}[t]
\centering
\includegraphics[width=\textwidth]{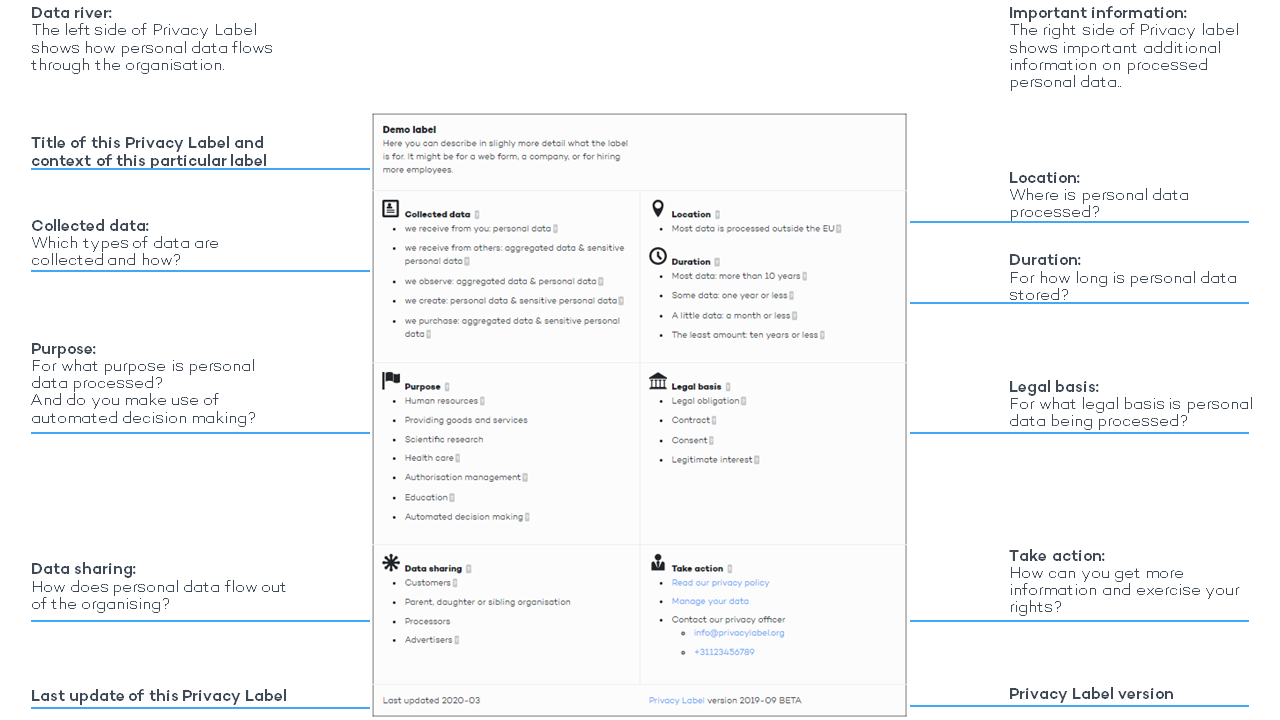}
\caption{Basic elements of the PrivacyLabel.org (NL.PL).}\label{fig_NL.PL_elements}
\end{figure}

The standardized model makes it easier to detect differences between labels of different service providers (e.g., \refGoal{goal_procurement}).
Take, for example, the case of how one can drill down for location information in the NL.PL. There is an icon for location saying that most data is processed outside the EU. One can click on a question mark for more information which says that the laws of other countries may apply, with some extra information and a link to learn more about what does location mean and what legal requirements apply. The basic information is about the location of the storage of data, if it is within the EU, `Yes' or `No'. If it is not, you have extra information on what does it mean not being stored in the EU, what kind of other different laws may apply and then you can click through it to learn even more. This is the way the NL.PL label is built around all the different main requirements the controllers have to inform data subjects about. It is a layered approach, where what you see on the label is some part of the information and there is the question mark after each sentence, where one can click and have some basic additional information and then you can click further to learn, reaching a knowledge base where one finds more information about what it actually means.
}

\section{Three Approaches to Managing PL}\label{sec_approaches}

\Multy%
Arriving to a PL with the characteristics and the goals described so far require a process involving all the stakeholders described in Section~\ref{sec_PL_Stakeholders}. Introducing and developing PL requires research efforts on the open problems that we have identified, but the adoption of PL requires involvement of more than the research communities. For the adoption to be successful, PL needs to show not only that it can solve problems of all these stakeholders, but also that it can live and thrive after the initial starting phase. This requires good management of PL, which could involve three different important aspects, all fitting together.
\begin{characteristic}
The privacy label can be a way to visually present and structure the privacy aspects detailed in a privacy policy, thus helping to implement the very important GDPR principle of transparency. This is typical of a self-evaluation approach, as taken by NL.PL, and detailed more in Section~\ref{subsec_approach_self}\longPaper{, but still tied to a legally binding document so that it cannot become a means of deceit}.
\end{characteristic}

\begin{characteristic}
The privacy label can be a means of establishing trust, usually created through an evaluation and audit process by a certification body, such as EuroPriSe, based on technical requirements that strive to identify whether existing legislations, such as the GDPR in Europe, are respected (see Section~\ref{subsec_approach_cert}).
\end{characteristic}

\begin{characteristic}
The privacy label can be a way to measure (the usability of) privacy on scales, allowing for comparisons. Measuring can be done either automatically (see Section~\ref{sec_automation}) or with the help of a community (see Section~\ref{subsec_approach_crowd}).
\end{characteristic}

\subsection{Self-evaluation}\label{subsec_approach_self}

\startingPoint{}
for doing self-evaluations of privacy and producing a privacy label is the quite advanced web-platform of PrivacyLabel.org (NL.PL). 
This combines a visual design that uses icons, with succinct textual descriptions, into an accessible way of presenting how an organization manages privacy. This is a service 
\longPaper{for doing self-evaluations, and as such, the }\onlyShortPaper{where an }%
organization by itself will have to explain and include information about the privacy measures taken, be that technical, PETs, legal, procedural, etc.
\longPaper{The self-evaluation can be done by the organization itself or perhaps with support from outside privacy experts. This is different from a trust mark or seal that imply an independent evaluation done by designated bodies, who are evaluating and then creating the label for the organization.}%
NL.PL focuses on transparency, allowing an organization to show that they are taking privacy seriously and how they are doing that.

\Reggy%
There is a crucial difference between internal auditing, as above, and external certifications that provide a seal issued by certification bodies, as referred in the Art.~42/43 of GDPR. This type of label is usually not covering the entire organization's privacy attitude. 
\longPaper{An external auditor first checks a specific system and its processing activities for compliance and then provides the label.}%
The NL.PL approach is to have a label that is in the style of nutrition facts labels, \longPaper{showing to consumers and customers more explanations about what the organization is doing,} making more of a summary of a privacy statement, i.e., aiming to make the privacy statement comprehensible for the customer\longPaper{ or the user of the service}. 
\longPaper{In a privacy mark there is more focus on the technical audit, while in the NL.PL there are a number of icons representing main categories such as whether the data is stored within the EU or outside, what data is used for, how long the retention terms are. 
The NL.PL contains basic information answering to basic requirements that are mandatory according to GDPR. It is more about the information duties than the technical official audit.}

\Multy%
Checking the validity of NL.PL can be done by either an external audit (see Section~\ref{subsec_approach_cert}), an automated process (see Section~\ref{sec_automation}), or by a community effort (see Section~\ref{subsec_approach_crowd}). 
NL.PL only focuses on the privacy statement replacement and transparency, and not on trust marks or seals, which are certificates connected to a specific processing activity, or a specific system. 
\Reggy%
A trust mark or privacy seal that involves an external audit is not a replacement for a privacy statement. The organization needs to have this anyway, and then NL.PL offers a form of the privacy statement that is more \longPaper{transparent and }comprehensible for the consumer. 

\Lancey%
Self-assessment can imply the assumption of honest controllers. However, one can think of methods to oversee that self-assessed PL do not become a means of deceit, e.g., by involving administrator fines, since being transparent is still enforced through Art.~12-13 of GDPR. 
\Bussy%
Transparency is also mandatory for the Common Market\longPaper{\footnote{The European Single Market, Internal Market or Common Market is a single market which seeks to guarantee the free movement of goods, capital, services, and labor -- the `four freedoms' -- within the European Union.}}.
\longPaper{The \CEsign\   marking\longPaper{\footnote{The \CEsign\ marking is a certification mark that indicates conformity with health, safety, and environmental protection standards for products sold within the European Economic Area.}} is an example of an existing self-assessment that is mandatory for more critical areas such as medical devices, which is a small fraction of the market. In this area it is mandatory to have the \CEsign\  marking even before getting the product on the market. This is an example of a self-assessment that does not involve an external expert looking over the manufacturing places.}%
A well known example of self-assessment comes from the e-waste management area \citep{ewaste2020CACM} known as EPEAT\footnote{Electronic Product Environmental Assessment Tool: \url{https://epeat.net/}} from the Green Electronics Council\longPaper{\footnote{\url{https://greenelectronicscouncil.org/epeat-criteria/}}}, where the participation is voluntary, yet over the years it has become quite adopted by manufacturers. 
\longPaper{The program provides labeling for electronic products that meet certain criteria across a range of 12 categories, covering materials and chemical usage, energy efficiency, recyclability, product lifespan, and product design.}

\begin{goal}
We do not expect PL to be something mandatory in a first phase, \longPaper{in the sense that external entities are checking for compliance, }but instead aiming first to have such self-declarations more widely spread.
\end{goal}

\subsection{Certification and Audit}\label{subsec_approach_cert}

\Reggy%
DPAs can accredit companies to do privacy certification (Art.~43 of GDPR); in Europe one of the most advanced is EuroPriSe which originated in 2001 from the German Schleswig-Holstein DPA.
Currently, audit and certification usually focuses on a system (or more often on one component that is considered critical for the system's security and privacy), and therefore it is important in a PL to properly identify the 
\emph{target of the evaluation}. As companies tend to put labels on the package, they might misinform, e.g., the ULD seal ended up on the boxes of the whole product, even though it concerned only the activation part. 
\longPaper{They had though a footnote mentioning that only the online validation tool received such a privacy seal.}

\Lancey%
Privacy agreements are important documents for providing information on data processing, as required by regulations all over the world.%
\footnote{GDPR in Europe, California Consumer Privacy Act, Personal Information Protection and Electronic Documents Act in Canada, Privacy Act 1988 in Australia \citep{2018AustraliaPrivacyLaw}.} 
It is common for lawyers to work with businesses to prepare and present privacy policies for certification or compliance audit; in which case devising a corresponding PL could be seen both as a certification seal as well as an explanatory label.%
\longPaper{Depending on the company and the target of evaluation, one could end up with very long agreements that are complicated and not easy to understand, in which case one is interested in presenting the information in a layered manner, simplifying and organizing the information so it becomes easier to be read by the consumers.}

\begin{characteristic}
For PL to be part of a certification scheme it needs a standardized way to present the information, harmonized to be suited for the multitude of actors that have to present rather diverse privacy aspects. 
\end{characteristic}

\longPaper{This is already the case with privacy policies, some using legal terminology, some simplifying it to terms understandable by a more general reader. }%
\longPaper{Depending on how a company wants to be perceived by their customers, they can work with their privacy policies to, e.g., simplify the language or highlight some of the important aspects, but also add videos, illustrations, symbols, or a combination of these as the PL. 
Sometimes it can be that a company (for some of their services) wants to be sure that the legal aspects are thoroughly covered, thus devising a longer, more complicated privacy policy. Other business strategies are more concerned with giving a good impression, in terms of having easily readable policies, with simpler text that is more easily understandable by the end users. 
}

\subsection{Crowd-sourced and Community driven}\label{subsec_approach_crowd}

There are numerous community efforts, from the wide\-spread open-source software developments (OSM) or the Wikimedia projects, to the more recent and relevant 
LeDA\footnote{Legal Design Alliance \url{https://www.legaldesignalliance.org}}
or 
ToS;DR\footnote{``Term of Service; Didn't Read'' community effort on explaining ToS \url{https://tosdr.org}} .
In the same spirit and management style, we envision a Community Coordinated Privacy Labeling, or CoCoPL
(see also the more general socio-technical framework called CoCoAI \citep{sivesind2021MSc}).
CoCoPL would include as part of the community both lawyers, e.g., from ToS;DR, as well as developers, e.g., from OSM projects, but also members from all the stakeholders identified in Section~\ref{sec_PL_Stakeholders}. One example is to involve the laypeople in a crowd-sourcing effort of annotating privacy policies to help the AI-based automation tools of Section~\ref{subsec_AI_automation}. Another example can be to involve the DPAs as more trusted members of the community, though a trust-model is first needed, with such communities usually employing meritocracy.
Internet communities have for a long time organized themselves in forums to evaluate businesses and products. Companies are well aware of this, and often misinformation becomes a problem that forums (or tech-magazines) have to deal with.
CoCoPL would do a similar activity of evaluating the privacy practices of businesses and products, to the benefit of the community and everyone else as already mentioned.

Crowd-sourcing the data annotation means that we can have continued expansion of the data sets. Engaging interest groups that have experience with privacy policies, such as ToS;DR, would help fine-grain the privacy concepts used in the annotation models and how these would refine the constructed PL. 
Another benefit of crowd-sourcing data labeling is easy adaptation to changing standards and trends.

\section{Concluding with a Timeline for Finding PL}\label{sec_roadmap}

Before concluding we devise an action plan for finding PL.

\vspace{1ex}\noindent\textbf{Phase I:} 
Attract companies and organizations to perform self-assessments, based on their own privacy policies, using a structuring tool such as the NL.PL. 

\vspace{1ex}\noindent\textbf{Phase II:} 
Bring in more of the automation and reasoning tools for creating privacy measurements, to identify the level of privacy protection to be shown by PL. 
Include psychological models to individualize PL, and give also a proper appearance for the intended purposes, one of these being educational.
Aim more on generating the PL automatically from reliable documents like the ToS and from inputs such as those from a user's privacy profile.

\vspace{1ex}\noindent\textbf{Phase III:} 
Join forces with the authorities for introducing a regulatory framework to make the self-assessment mandatory and uniform.
At the same time, build a community around PL, that could even include the authorities\longPaper{ as one (rather important) member in the community}.

\vspace{2ex}
\Reggy
The self-evaluation is a logical first step from a GDPR perspective, because of the large emphasis on demonstrating compliance. The controllers need to show how they abide by the GDPR and how they implement the data protection principles. 
Further on, one can introduce checks and fines, including external audits and certifications, done by the authorities or the community.

\subsection*{Concluding remarks}

When dealing with a complicated concept such as privacy, that is faced with multiple long-standing problems as discussed in Section~\ref{sec_sleeping_princess_privacy}, then to develop a solution can be a daunting task. 
In response, we propose an all-encompassing definition of Privacy Labeling as a possible start on the road towards a solution to many of the current privacy problems our society is struggling with.

When developing such a panoptic concept as the Privacy Labels proposed here, it is a good idea to find many discussion partners among the various stakeholder groups. Therefore, the ideas that we have presented have roots in our conversations with experts from the seven different fields that we considered relevant for privacy labeling. Our goal with bringing all these views was to investigate the concept of privacy labeling (and its implications) from many different angles. No one single discussion partner has `the right answer', but their collective opinions sum up to a result that is much more comprehensive and powerful than any one view could accomplish on its own.

\notAnonymousPaper{
\onlyShortPaper{
\subsection*{Acknowledgments}
We would like to thank associate Torunn Hellvik Olsen for her great inputs during our workshop on this topic held in Oslo, March 2020.
}%
}


%

\end{document}